\documentclass[aps,prb, twocolumn, superscriptaddress, amsmath, amsfonts, amssymb, amsthm]{revtex4-2}
\usepackage{graphicx}
\usepackage{dcolumn}
\usepackage[unicode=true,
 bookmarks=false,
 breaklinks=true,pdfborder={0 0 1},backref=false,colorlinks=true]
 {hyperref}
\hypersetup{pdfcreator={},pdfproducer={LaTeX with hyperref},linkcolor=blue,anchorcolor=blue,citecolor=blue,filecolor=red,menucolor=red,urlcolor=blue,pdfstartview=FitV,pdfhighlight=/I,hypertexnames=true}
\makeatletter
\usepackage{lmodern}
\usepackage{mathptmx, newtxtext, newtxmath, xspace}
\usepackage{colortbl}
\usepackage{array}
\usepackage{bbm}
\usepackage{siunitx}
\usepackage{ulem}
\usepackage{comment}
\usepackage[table, dvipsnames]{xcolor}
\newcommand{\hhbar}{\mathchar'26\mkern-9mu h}

\usepackage{bm}
\usepackage{braket}

\definecolor{colorA}{cmyk}{0,0,0,0.05}
\definecolor{colorB}{cmyk}{0.14,0.04,0,0}
\definecolor{colorC}{cmyk}{0.02,0.0799,0,0}
\definecolor{colorD}{cmyk}{0.099,0.14,0,0}
\definecolor{FMirrepColour}{gray}{0.4}

\usepackage{natbib}
\setcitestyle{numbers}

\makeatother

\begin{document}
\title{Elasto-Hall conductivity and the anomalous Hall effect in altermagnets}
\author{Keigo Takahashi }
\affiliation{Department of Physics, University of Tokyo, Bunkyo, Tokyo 113-0033, Japan}
\author{Charles R. W. Steward}
\affiliation{Institute for Theory of Condensed Matter, Karlsruhe Institute of Technology,
Karlsruhe 76131, Germany}
\author{Masao Ogata }
\affiliation{Department of Physics, University of Tokyo, Bunkyo, Tokyo 113-0033, Japan}
\affiliation{Trans-Scale Quantum Science Institute, University of Tokyo, Bunkyo, Tokyo 113-0033, Japan}
\author{Rafael M. Fernandes}
\affiliation{Department of Physics, The Grainger College of Engineering, University of Illinois Urbana-Champaign, Urbana, IL 61801, USA}
\affiliation{Anthony J. Leggett Institute for Condensed Matter Theory, The Grainger College of Engineering, University of Illinois Urbana-Champaign, Urbana, IL 61801, USA}
\author{J\"org Schmalian }
\affiliation{Institute for Theory of Condensed Matter, Karlsruhe Institute of Technology,
Karlsruhe 76131, Germany}
\affiliation{Institute for Quantum Materials and Technologies, Karlsruhe Institute
of Technology, Karlsruhe 76131, Germany}
\date{\today }
\begin{abstract}
Altermagnets break time-reversal symmetry, preserve the crystal translation invariance, and have a spin density with $d$-wave, $g$-wave, etc. momentum dependencies which do not contribute to the magnetization. 
When an $s$-wave spin-density contribution cannot be excluded by symmetry a small magnetization and  an anomalous Hall effect (AHE) emerge. However, for so-called "pure" altermagnets, where the $s$-wave component is symmetry forbidden even in the presence of SOC, both the zero-field magnetization and the AHE vanish. We show that altermagnets generally exhibit a non-zero elasto-Hall-conductivity, by which application of strain leads to a non-zero AHE. For pure altermagnets it is the only contribution to the AHE. This elasto-Hall-conductivity is caused by strain coupling to the Berry curvature quadrupole that characterizes altermagnets and allows for the determination of the altermagnetic order using transport measurements that are linear in the electrical field. We further show  that the emergence of a non-zero magnetization in the presence of strain arises from a different response function: piezomagnetism. While this magnetization gives rise to an additional contribution to the elasto-Hall conductivity, the corresponding Berry curvature is qualitatively different from the distorted Berry curvature quadrupole originating from the altermagnetic order parameter. 
This insight also helps to disentangle  AHE and weak ferromagnetism for systems with symmetry-allowed  $s$-wave contribution.
Quantitatively, the elasto-Hall conductivity is particularly pronounced for systems with a Dirac spectrum in the altermagnetic state. The same mechanism gives rise to anomalous  elasto-thermal Hall,  Nernst,  and Ettinghausen effects.

\end{abstract}
\maketitle

\section{Introduction}
Altermagnets are magnetically ordered systems that, unlike ferromagnets, possess - at least in the ideal or "pure" case - no net magnetization and, unlike antiferromagnets, do not break the crystal translation invariance \citep{Smejkal2022,Smejkal2022b,vsmejkal2020crystal}. Instead, time-reversal symmetry and a point group operation other than inversion are spontaneously broken, while their product continues to be a symmetry (for a recent review, see \cite{Jungwirth2024altermagnets}).
This enforces the magnetic moments of the atoms at different (but symmetry-related) crystallographic positions to cancel each other, resulting in a magnetization that averages to zero. Thus, while a multipole expansion of the magnetization density yields a vanishing dipole moment, the spin density in momentum space acquires a $d$-wave, $g$-wave, or $i$-wave form factor, resulting in spin-split bands even in the absence of spin-orbit coupling (SOC) \citep{Smejkal2022}. 
States now known as altermagnets are closely related to correlated-electronic phenomena such as multipolar magnetism \cite{voleti2020multipolar,Hayami2019,fiore2022modeling,Bhowal2022,Urru2022,winkler2023theory} and Pomeranchuk instabilities of higher angular momentum in the spin channel \cite{Pomeranchuk1958,Fradkin2007,Jungwirth2024supefluid}, spin-valley locked antiferromagnets~\cite{ma2021multifunctional,hu2024catalogue}, and  order proposed for  the organic charge transfer salts \cite{Naka2019,Naka2020}.
A flurry of experimental and theoretical research has resulted in the identification of unique altermagnetic phenomena \citep{mazin2021prediction,Smejkal2022,Smejkal2022b,Turek2022,Bhowal2022,vsmejkal2020crystal,Smejkal2022chiral,betancourt2023spontaneous,mazin2023altermagnetism,Autieri2023_perovskites,Mazin2023_FeSe,Chakraborty2023,feng2022anomalous,bai2023efficient,Steward2023,Facio2023,Ding2024large,Yang2024three,Li2024topological,fernandes2024topological,mcclarty2024landau,Liu2024,Fang2024,Song2024,Ghorashi2024,Sodequist2024,giil2024superconductor,hodt2024spin,jaeschke2024supercell,Haule2024,krempasky2024altermagnetic,Amin2024nanoscale} and minimal models for altermagnetism \cite{Sudbo2023,Antonenko2024,Agterberg2024,SchnyderTopology,Attias2024,Knolle2024}.

 In some cases a spin density which is dominated by $d$-wave, $g$-wave etc. contributions may  allow, by symmetry,  an $s$-wave part and the
system establishes a finite zero-field magnetization in the presence of SOC. Such an altermagnet may have rather different electronic properties, but breaks the same symmetry as a ferromagnet. The primary focus of this paper, which always assumes a finite spin-orbit coupling, are pure altermagnets \cite{fernandes2024topological} where the $s$-wave part and hence the magnetization are forbidden by symmetry.
Given the absence of a net magnetization, the order parameter of a pure altermagnetic state does not directly couple to an external magnetic field, which complicates its experimental determination. It is therefore desirable to identify experimental methods that can be utilized to determine whether a given system possesses a finite altermagnetic order parameter.
Ideally, one wants to identify a transport or thermodynamic experiment that accomplishes this. 


\begin{figure*}
 \centering
 \includegraphics[width=\linewidth]{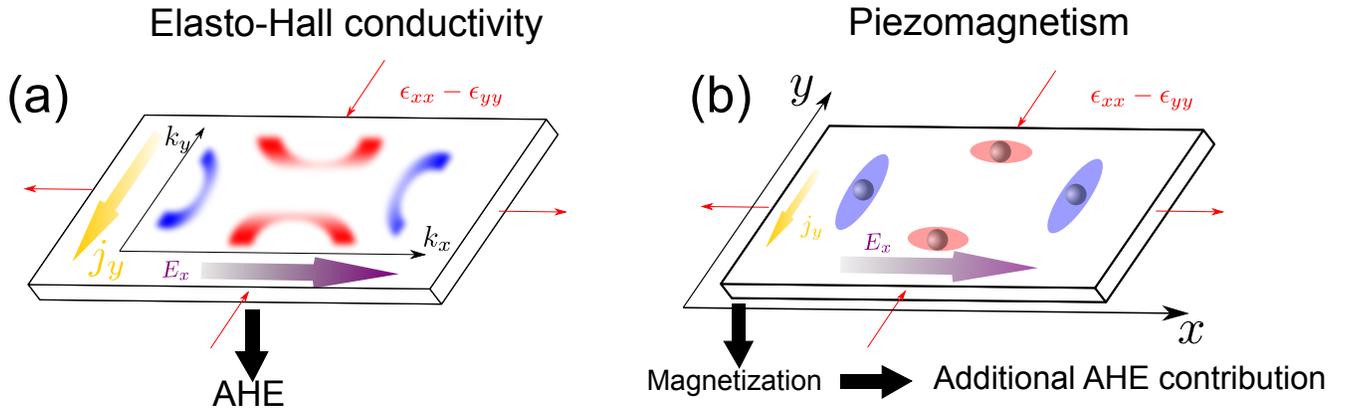}
 \caption{The anomalous Hall conductivity in strained altermagnets has two contributions: One from the change in the Berry curvature quadrupole under strain, associated with the response function called elasto-Hall conductivity (panel (a)); the other one from the strain-induced magnetization, associated with a distinct response function called piezomagnetism (panel (b)). Symmetry constraints enforce the two responses to be either both zero or both non-zero. The quadrupolar form of the Berry curvature with blue for positive values and red for negative values is shown in panel (a). For the unstrained system,  it leads to no AHE as the Berry curvature averages to zero over the Brillouin zone. Straining the system  leads to the formation of a net Berry curvature monopole which is manifested in the appearance of an AHE. The other (typically smaller) contribution comes from the net magnetization induced via the piezomagnetic effect that couples the altermagnetic order parameter with magnetization and symmetry-breaking strain (panel (b)).  The net magnetization arises from the deformation, by strain, of the spin density around each sublattice in the unit cell (blue for spin up and red for spin down), as shown in panel (b). 
 The current $j_y$ shown in the two panels is linear in the electric field $E_x$ and in the strain $\epsilon_{xx}-\epsilon_{yy}$, giving rise to the non-zero elasto-Hall conductivity elements $\nu^{(a)}_{xyxx}-\nu^{(a)}_{xyyy}\neq 0$, as defined in Eq.~\eqref{eq:elasto_cond}, and the non-zero piezomagnetic tensor components $\Lambda_{zxx} - \Lambda_{zyy} \neq 0$, as defined in Eq. (\ref{eq:magnetization}). The smaller current $j_y$ in panel (b) illustrates that the AHE contribution arising from the magnetization is typically smaller than that arising from the effect in panel (a).}
 \label{fig:magnetisation}
\end{figure*}

The possibility of intrinsic thermal or electrical Hall transport in nominally compensated magnets was discussed for a number of non-trivial non-collinear and collinear magnetic-moment configurations \cite{MacDonald2014,Naka2019, Naka2020, Hayami2019, Naka2022,vsmejkal2022anomalous}, including those associated with altermagnetism \cite{vsmejkal2020crystal,Attias2024, Zhou2024}. 
 In the case of altermagnets, the AHE can only emerge for those orientations of the magnetic moment and in the presence of SOC, which also allow for the emergence of weak ferromagnetism with an $s$-wave component of the spin density \cite{vsmejkal2020crystal,Smejkal2022b,fernandes2024topological}. 
These are manifestations of a general symmetry constraint that the AHE and a net dipole moment are either both present or both absent. Thus, in the examples above, despite the fact that the starting magnetic configurations with finite AHE are nominally compensated, this compensation is not symmetry enforced,
 resulting in a possibly small but finite net ferromagnetic moment. Note, this relationship does not imply \textit{causality}, i.e., one cannot conclude that the AHE is caused by the allowed ferromagnetic moment and vice versa. While these considerations are relevant for systems with symmetry-allowed $s$-wave component,  there is no spontaneous anomalous Hall effect in pure altermagnets.

Although the AHE is not a generic property of altermagnetism, the proclivity of altermagnets with an $s$-wave component to display a sizable AHE suggests the existence of a ``hidden'' related response even in pure altermagnets \cite{fernandes2024topological}, i.e. altermagnets that do not induce a weak magnetization even in the presence of SOC. To shed light on this issue, we note that the intrinsic Hall effect results from the Berry curvature, an axial vector that acts like a magnetic field in momentum space, deflecting electrons transversely \cite{Karplus1954,Chang1996,Sundaram1999,Haldane2004,Nagaosa2010}. In altermagnets, the Berry curvature has a quadrupolar or higher multipolar symmetry in momentum space, yielding a vanishing averaged Berry curvature and hence no anomalous Hall effect \cite{vsmejkal2020crystal,Antonenko2024,Cano2024,Knolle2024,Sorn2024,Farajollahpour2024}. The situation is analogous to time-reversal symmetric systems with broken inversion symmetry, where the Berry curvature forms a dipole in momentum space. In that case, however, a non-zero Hall effect can be induced when, through nonlinear electric field processes, an electronic distribution function emerges that differs for states with opposite sign of the Berry curvature \cite{Sodemann2015}. As a result, one obtains a non-linear anomalous Hall effect induced by the Berry curvature dipole, an effect that occurs at second order in the electrical field. 

Generalizations of this approach have been made to yield non-linear Hall effects at third order in the field for compensated magnets \cite{Xiang2023,Sankar2024} and altermagnets \cite{Sorn2024,Farajollahpour2024,Cano2024}, described by
$ j^{(3)}_{\alpha}=\sigma^{(a)}_{\alpha\beta\gamma\delta}E_{\beta}E_{\delta}E_{\gamma} $. Here the sum over repeated indices is implied, $j^{(3)}_\alpha$ denotes a current component at third order, $E_\alpha$ denotes an electric field component, and the superscript $(a)$ denotes the anti-symmetric part of the conductivity (with respect to the first index pair). The Berry curvature quadrupole requires the additional power in the electrical field compared to the Berry curvature dipole of Ref.~\cite{Sodemann2015}. This yields a transverse current that is third-order in the field.  While these non-linear transport effects are interesting, in particular in the ac regime, there is a question that is both of practical and conceptual interest, since higher-order responses in the field can be very small and difficult to measure: is there a transport coefficient that is linear in the electric field and that gives an anomalous Hall current as a unique fingerprint of the Berry curvature quadrupole in an altermagnet?

To answer this question, we note that altermagnets are invariant under a combination of time-reversal symmetry and rotations, and that rotational symmetry can be broken by uniaxial strain along specific directions. Thus, the explicit breaking of rotational symmetry by strain in an altermagnet should result in the breaking of time-reversal symmetry alone, thus leading to a non-zero Berry curvature and to an AHE. Therefore, we focus on the elasto-conductivity (also known as piezo-conductivity) $\nu_{\alpha\beta\gamma\delta}$, which is defined through
\begin{equation}
 j_{\alpha}=\sigma_{\alpha \beta} E_\beta+\nu_{\alpha\beta\gamma\delta}E_{\beta}\epsilon_{\gamma \delta},
 \label{eq:elasto_cond_gen}
\end{equation}
where $\epsilon_{\gamma \delta}$ is the strain tensor. Hence, the second term describes the impact of strain variations on changes in the electrical transport. The analysis of the elasto-transport has been used in the investigation of semiconductors \cite{Sun2009} and played a crucial role in identifying the role of nematic fluctuations and order in iron-based superconductors~\cite{Kuo2013}, as well as for a range of exotic properties in charge density wave states \cite{Nie2022,Frachet2022,Singh2024}. The power of the elasto-transport analysis lies in its ability to slightly change the crystalline symmetry of a material, imposing  constraints on the permissible states of electronic order \cite{Shapiro2015,Sorensen2021}. Indeed, it has been recently shown to be a fingerprint of another unusual ordered state called ferroaxial order \cite{DayRoberts2025}.

In this paper, we show that altermagnets intrinsically possess a non-zero elasto-Hall conductivity response, defined as the tensor components $\nu^{(a)}_{\alpha\beta\gamma\delta}$ of $\nu_{\alpha\beta\gamma\delta}$ that are {\it antisymmetric} in the first two indices. This response gives rise to a Hall effect that is linear in the electrical field. More formally, since the Hall components of the conductivity at zero strain, i.e. $\sigma^{(a)}_{\alpha\beta}$, vanish by symmetry in pure altermagnets, it holds
\begin{equation}
j_{\alpha}=\nu^{(a)}_{\alpha\beta\gamma\delta}E_{\beta}\epsilon_{\gamma \delta}\, .
 \label{eq:elasto_cond}
\end{equation}
Microscopically, this response arises because straining the system distorts the Berry curvature quadrupole of an altermagnet and induces a finite Berry curvature monopole to leading order in strain, giving rise to a Hall voltage. We first demonstrate the effect for a specific minimal model of a tetragonal system with a $d_{x^2-y^2}$ altermagnetic order parameter and then offer a detailed analysis of the symmetry-allowed anomalous Hall elasto-conductivities for generic trigonal, tetragonal, hexagonal, cubic, and orthorhombic altermagnets. Except for one even-parity, time-reversal-odd order parameter symmetry in cubic systems, all of these crystalline point groups allow for finite elements of $\nu^{(a)}_{\alpha\beta\gamma\delta}$ that can be used to identify the underlying altermagnetic order experimentally. While the existence of the elasto-Hall conductivity was previously mentioned in Ref.~\cite{hu2024catalogue}, here we explore in depth the microscopic mechanisms and its connection to piezomagnetism.

The application of certain strain components on an altermagnet is also known to endow it with a non-zero magnetization via the piezomagnetic effect \cite{ma2021multifunctional,Steward2023,fernandes2024topological,mcclarty2024landau,vandenBrink2024}:

\begin{equation}
 M_{\mu}= \Lambda_{\mu \gamma \delta} \epsilon_{\gamma \delta}
 \label{eq:magnetization}
\end{equation}
where $M_{\mu}$ is the $\mu$-th component of the magnetization density  and  $\Lambda_{\mu \gamma \delta}$ is the piezomagnetic tensor. We use group theory to show that the two tensors, namely, the elasto-Hall conductivity $\nu^{(a)}_{\alpha\beta\gamma\delta}$ and the piezomagnetism $\Lambda_{\mu \gamma \delta}$ (more precisely $\varepsilon_{\alpha \beta \mu}\Lambda_{\mu \gamma \delta}$), have the same irreducible representation decomposition in any point group. As a result, if an element of one of the tensors is non-zero, an element of the other tensor must also be non-zero. Conversely, if all elements of one of the tensor are zero, then all elements of the other tensor are also zero. This is a direct consequence from the fact that the Hall vector and the magnetization have the same transformation properties. 

Using the same minimal model of a $d_{x^2-y^2}$ altermagnet, we compute the piezomagnetic tensor $\Lambda_{\mu \gamma \delta}$ microscopically. Comparison with the expression of the elasto-Hall conductivity $\nu^{(a)}_{\alpha\beta\gamma\delta}$ confirms that microscopically, different processes contribute to these response functions. Of course, a non-zero magnetization also generates another contribution to the AHE on top of the AHE generated via the elasto-Hall conductivity. We show that the Berry curvature density associated with the magnetization-induced AHE has a qualitatively different momentum dependence than the distorted Berry curvature quadrupole arising from the altermagnetic order parameter. This further demonstrates that the AHE in altermagnets is not a consequence of the non-zero magnetization. The existence of two independent response functions in altermagnets, elasto-Hall conductivity and piezomagnetism, is illustrated in Fig. \ref{fig:magnetisation}.

This paper is organized as follows. Section \ref{sec:Berry_AM} presents a calculation of the Berry curvature density in a Lieb-lattice low-energy model for $d_{x^2-y^2}$ altermagnetism. In Section \ref{sec:elasto_Hall}, we compute the elasto-Hall conductivity of this model and extend the results to the thermoelectric and thermal analogs of the elasto-Hall conductivity. Section \ref{sec:piezomagnetism} computes the piezomagnetic response of the low-energy model. In Section \ref{sec:symmetry}, group theory is employed to determine the non-zero components of the elasto-Hall conductivity tensor in common point groups, as well as establish its relationship to the piezomagnetic tensor. Section \ref{sec:conclusions} is devoted to the conclusions. Finally, Appendix~\ref{Appendix_Dirac} presents an analytic approximation for the elasto-Hall conductivity of the low-energy model.

\section{Berry curvature quadrupole in altermagnets}  \label{sec:Berry_AM}

\subsection{Hall conductivity and Berry curvature}
Before we discuss the elasto-Hall conductivity, we summarize some of the main ingredients of the intrinsic Hall conductivity $\sigma^{(a)}_{\alpha \beta}$ that occurs as a result of a finite Berry curvature \cite{Karplus1954,Chang1996,Sundaram1999,Haldane2004,Nagaosa2010}. It is given by
\begin{equation}
 \sigma^{(a)}_{\alpha \beta}=-\frac{e^2}{\hhbar }\sum_{\bm{k}\in {\rm BZ},~l}f\left(\xi_{\bm{k},l}\right)\Omega_{\alpha \beta}^{(l)}\left(\bm{k}\right),
 \label{eq:Hall_cond}
\end{equation}
with $\alpha \neq \beta$, Fermi function $f\left(\xi_{\bm{k},l}\right)$, and $\xi_{\bm{k},l}$ the energy eigenvalue for band $l$ and crystal momentum $\bm{k}$ with Bloch states $\ket{u_{\bm{k}, l}}$. $\Omega^{(l)}_{\alpha\beta}(\bm{k})$ is the Berry curvature of the $l$-th band and given by
\begin{align}
\Omega_{\alpha \beta}^{(l)}\left(\bm{k}\right) = i \left[ \Braket{\frac{\partial u_{\bm{k},l}}{\partial k_{\alpha}}| \frac{\partial u_{\bm{k},l}}{\partial k_{\beta}}} - [\alpha \leftrightarrow \beta] \right].
\label{eq:BerryC}
\end{align}
It is convenient to introduce the frequency dependent function
\begin{align}
\sigma^{(a)}_{\alpha \beta}(\omega) = - \frac{e^2}{\hhbar} \sum_{\bm{k} \in \text{BZ},~l} \Omega^{(l)}_{\alpha \beta}(\bm{k}) \Theta(\omega - \xi_{\bm{k},l}),
\label{eq:sigma_omega}
\end{align}
where $\Theta(x)$ is the Heaviside function. Then it holds
\begin{equation}
 \sigma^{(a)}_{\alpha \beta}=\int d\omega \left(-\frac{\partial f\left(\omega\right)}{\partial \omega}\right) \sigma^{(a)}_{\alpha \beta}(\omega). 
 \label{eq:Hall_cond_freq}
\end{equation}
Note, $\sigma^{(a)}_{\alpha \beta}(\omega)$ does not correspond to an optical conductivity but is rather an efficient quantity to determine  d.c. conductivity. The anomalous d.c. Hall conductivity at zero-temperature is given by $\sigma^{(a)}_{\alpha \beta}(E_{\rm F}) $, with Fermi energy $E_{\rm F}$.

For our subsequent analysis the transformation properties of the Berry curvature will play an important role. To this end we consider the three-component vector 
\begin{equation}\boldsymbol{\Omega}\left(\boldsymbol{k}\right)=\left(\Omega_{yz}\left(\boldsymbol{k}\right),\Omega_{zx}\left(\boldsymbol{k}\right),\Omega_{xy}\left(\boldsymbol{k}\right)\right), \label{eq:Berry_vector}
\end{equation}
which transforms under time reversal $\cal{T}$ and parity ${\cal P}$ like an axial vector that is time-reversal-odd
\begin{eqnarray}
\boldsymbol{\Omega}\left(\boldsymbol{k}\right) & \stackrel{\cal{T}}{\rightarrow} & -\boldsymbol{\Omega}\left(-\boldsymbol{k}\right),\nonumber \\
\boldsymbol{\Omega}\left(\boldsymbol{k}\right) & \stackrel{{\cal P}}{\rightarrow} & \boldsymbol{\Omega}\left(-\boldsymbol{k}\right) .
\end{eqnarray}
Hence, for systems with inversion and time-reversal symmetry $\boldsymbol{\Omega}\left(\boldsymbol{k}\right)=0$. The
 exceptions are isolated points with band degeneracies in the spectrum - giving rise to Dirac points - that serve as point sources for the Berry curvature; however, the Berry curvature has opposite signs for opposite momenta, such that $\sigma^{(a)}_{\alpha \beta}=0$. Interestingly, for systems with Dirac points, small perturbations that either introduce a small gap that breaks $\cal{T}$ and $\cal{P}$ or change the symmetry between momenta with opposite signs of $\boldsymbol{\Omega}\left(\boldsymbol{k}\right)$, give rise to large contributions to the Berry curvature. This last aspect plays an important quantitative role in our subsequent analysis.
 
 To obtain a finite Hall conductivity, the Berry curvature must of course be finite and should not vanish after averaging over momenta and the occupied bands. This is often called a finite Berry curvature monopole, which is the case for ferromagnets \cite{Karplus1954,Chang1996,Sundaram1999,Yao2004,Haldane2004,Nagaosa2010,Zhang2011}. Alternatively,  models with antiparallel magnetic moments were discussed where the symmetries of the Hamiltonian allows for a ferromagnetic moment. For example Naka {\it et al.} \cite{Naka2020} discussed a mechanism of doping-induced anomalous Hall effect (AHE) in a Hubbard model where there are four sites in the unit cell and also in an effective anisotropic Hubbard model with two sites in the unit cell, considering the two-dimensional $\kappa$-type organic compound \cite{Naka2019,Naka2020}.
 In systems with broken inversion symmetry but intact time-reversal symmetry, the Berry curvature can be finite, yet the monopole vanishes and only the dipole is nonzero. Hence, the linear Hall conductance also vanishes. 

\subsection{Vanishing anomalous Hall conductivity in unstrained altermagnets}
For an altermagnet, the leading non-zero multipole of the Berry curvature is a quadrupole \cite{vsmejkal2020crystal,Antonenko2024,Cano2024,Knolle2024,Sorn2024,Farajollahpour2024}. To see this, recall that in altermagnets the product ${\cal T R} $ of time reversal and a point-group element ${\cal R}$ that is not inversion keeps the magnetic state unchanged. This implies 
\begin{equation}
\boldsymbol{\Omega}\left(\boldsymbol{k}\right)=-\overline{R}\boldsymbol{\Omega}\left(-R^{-1}\boldsymbol{k}\right),
\label{eq:BC_inv_AM}
\end{equation}
where $R$ and $\overline{R}=R \det R$ are the representations of ${\cal R}$ for a
polar and axial vector, respectively. To see how this gives rise to a Berry curvature quadrupole, we first analyze the two-band model of Ref.~\cite{Antonenko2024} for a tetragonal $d_{x^2-y^2}$ altermagnet. This low-energy model \cite{Sudbo2023} captures many essential features of altermagnets, particularly in what concerns their topological properties \cite{Antonenko2024} and was discussed in the context of microscopic  sublattice interference leading to altermagnetism \cite{durrnagel2024altermagnetic}. Importantly, it also describes concrete tetragonal layered altermagnetic materials such as La$_2$O$_3$Mn$_2$Se$_2$ \cite{Wei2024_2} as well as the related compounds KV$_2$Se$_2$O \cite{Jiang2024discovery}, Rb$_{1-\delta}$V$_2$Te$_2$O \cite{Zhang2024crystal}, and V$_2$Te$_2$O \cite{Li2024strain}.

\subsubsection{Low-energy model}
\label{sec:two_band_model}

\begin{figure}[tbp]
\begin{center}
\rotatebox{0}{\includegraphics[angle=0,width=1\linewidth]{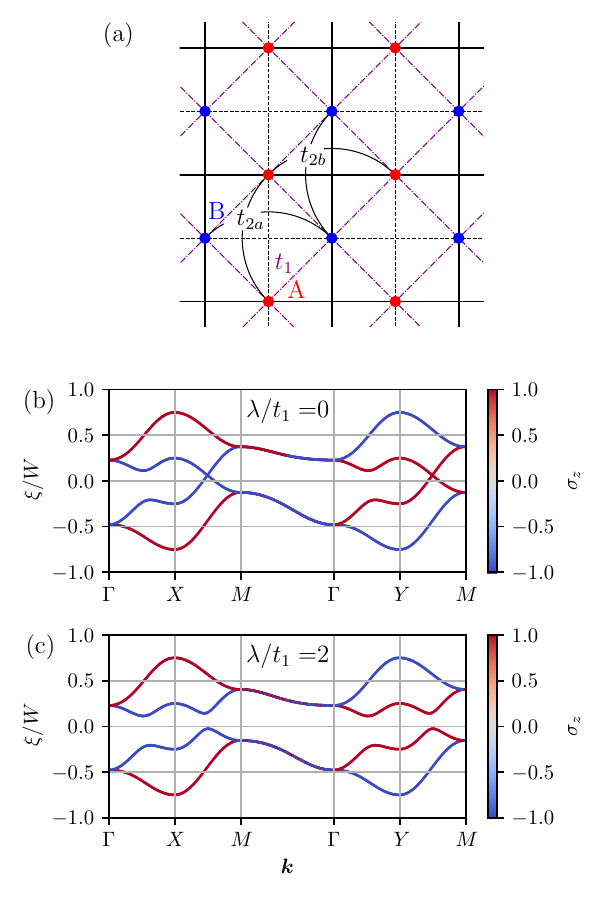}}
\caption{(a) Illustration of the Lieb lattice employed in  the low-energy model of electrons coupled to an altermagnetic order parameter (see Ref. \cite{Antonenko2024}). Panels (b) and (c) show the energy dispersions of the model described by Eq.~\eqref{eq:Hamiltonian}. We set $t_{2}/t_{1} = 0.5$, $t_{d}/t_{1} = 2$, and $N_{z} = 0.5 N_c$ with $N_c = |4t_{d}/J|$ \cite{Antonenko2024}. For the spin-orbit coupling, (b) $\lambda/t_{1} = 0$ and (c) $\lambda/t_{1} = 2$ are used. Energy is normalized by $W = 8t_{d}$. The colors represent the projection of the spin along the $z$-direction, with red indicating spin-up and blue indicating spin-down.}
\label{fig:band_no_strain}
\end{center}
\end{figure}

The low-energy model employed here consists of electrons on a Lieb lattice coupled to an altermagnetic order parameter, as illustrated in Fig. \ref{fig:band_no_strain}(a) \cite{Antonenko2024}. As such, electrons have both a sublattice and a spin degree of freedom. Because the two sublattices are related by a $90^\circ$ rotation, an intra-unit-cell antiparallel alignment of the spins results in a $d_{x^2-y^2}$ altermagnet with spin-split bands, except along the diagonals. The Hamiltonian is given by 
\begin{equation}
 H=\sum_{\boldsymbol{k}\in{\rm BZ}}c_{\boldsymbol{k}}^{\dagger}{\cal H}_{\boldsymbol{k}}c_{\boldsymbol{k}} ,
\end{equation}
where $c_{\boldsymbol{k}}=\left(c_{\boldsymbol{k},1},\ldots,c_{\boldsymbol{k},4}\right)^{T}$
combines the fermionic operators of the four flavor indices corresponding to
sublattice and spin degrees of freedom. The $4\times 4 $ Hamiltonian matrix 
\begin{equation}
{\cal H}_{\boldsymbol{k}}={\cal H}_{\boldsymbol{k}}^{0}+{\cal H}_{\boldsymbol{k}}^{N},
\label{eq:Ham_matrix}
\end{equation}
consists of a non-interacting part ${\cal H}_{\boldsymbol{k}}^{0}$ and the coupling of electrons to the altermagnetic order parameter ${\cal H}_{\boldsymbol{k}}^{N}$. In detail,
\begin{align}
{\cal H}_{\boldsymbol{k}}^{0}
=& - 4t_{1} \cos \frac{k_x}{2} \cos \frac{k_y}{2} \tau_{x} - 2t_2 (\cos k_x + \cos k_y) \tau_{0} \nonumber\\
&- 2t_d (\cos k_x - \cos k_y) \tau_{z} + \lambda \sin \frac{k_{x}}{2} \sin \frac{k_y}{2} \tau_{y} \sigma_{z}
\label{eq:Hamiltonian}
\end{align}
is the non-interacting Hamiltonian that includes the spin-orbit coupling (SOC) $\lambda$ term, nearest-neighbor hopping $t_1$ and anisotropic next-nearest-neighbor hopping elements $t_{2a}$ and $t_{2b}$, from which we define $t_2 = (t_{2a} + t_{2b})/2$ and $t_{d} = (t_{2a} - t_{2b})/2$.  $\tau_{\mu}$  and $\sigma_{\mu}$ are the Pauli matrices in sublattice and spin space, respectively,  where $\mu \in \{0,x,y,z\}$. In addition, we  have
\begin{eqnarray}
 {\cal H}_{\boldsymbol{k}}^{N}=J \tau_{z} N_{z} \sigma_{z},
 \label{eq:H_N}
\end{eqnarray}
which describes the coupling between the altermagnetic order parameter and the itinerant electrons. A natural scale of the order parameter is $N_c = |4t_{d}/J|$ \cite{Antonenko2024}. In the following, $J > 0$ is assumed. In terms of the irreducible representations of $D_{4h}$, $\tau_x$, $\tau_y$, and $\tau_z$  transform as $A_{1g}^+$, $B_{1g}^-$, and $B_{1g}^+$, respectively, while $\left(\sigma_x,\sigma_y\right)$ transform as $E_g^-$ and $\sigma_z$, as $A_{2g}^-$ (here, the superscript $\pm$ denotes whether the irrep. is even or odd under time-reversal, respectively). The altermagnetic order parameter $N_z$  transforms as $B_{2g}^{-}$, i.e. it is odd under time reversal and odd under rotations by $\pi/2$ about the $z$-axis or mirror reflections along the axes. Note that, because we include the SOC term, the magnetic moment direction has to be specified. For the out-of-plane moments chosen here, the altermagnetic order parameter does not induce weak ferromagnetism.

The energy dispersion at finite $N_z$ is shown in Fig.~\ref{fig:band_no_strain}, recovering the results reported in Ref. \cite{Antonenko2024}. The colors describe the projection of the spin along the $z$-direction, where red refers to spin up and blue to spin down. No spin splitting occurs along the high-symmetry line $\Gamma-M$, which is caused by the mirror symmetry with respect to the plane $k_y = k_x$. Panel (b) shows the dispersion for zero SOC, where one finds spin-polarized massless Dirac cones along $X-M$ and $Y-M$, provided that $|N_z| < N_c$, as discussed in Ref.~\cite{Antonenko2024}. The Dirac spectrum becomes massive once we include the SOC, as shown in Fig.~\ref{fig:band_no_strain}(c).

\begin{figure}[tbp]
\begin{center}
\rotatebox{0}{\includegraphics[angle=0,width=0.85\linewidth]{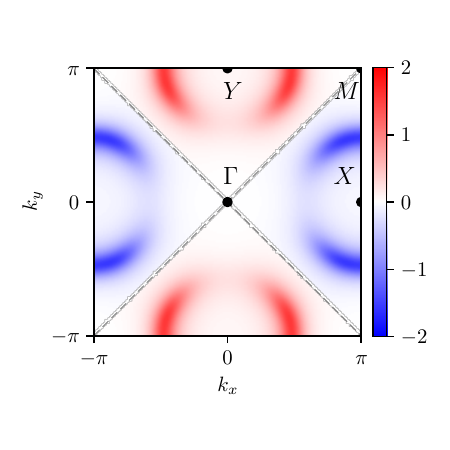}}
\caption{ The Berry curvature $\Omega_{xy}^{(2)}(\bm{k})$ of the second lowest band for the parameters used in Fig.~\ref{fig:band_no_strain}(b). The colors represent the sign of $\Omega_{xy}$, with red indicating positive and blue indicating negative values, i.e. one obtains a Berry curvature quadrupole where a rotation by $\pi/2$ changes the sign. The dotted lines indicate the nodal lines.}
\label{fig:Berry_no_strain}
\end{center}
\end{figure}
The Berry curvature of the second lowest band, $\Omega^{(2)}_{xy}(\bm{k})$, is shown in Fig.~\ref{fig:Berry_no_strain}, where the color indicates the sign of $\Omega^{(2)}_{xy}(\bm{k})$ (recall that the superscript $(2)$ refers to the band label). One notes that 
 a finite Berry curvature arises around the Dirac points and that the Berry curvature has a quadrupole structure with
\begin{equation}
\Omega^{(l)}_{xy}(k_x, k_y) =- \Omega^{(l)}_{xy}(-k_y, k_x),
\end{equation}
which is a special case of Eq.~\eqref{eq:BC_inv_AM}. This symmetry constraint leads, despite the large magnitude of the Berry curvature, to the absence of the anomalous Hall conductivity, i.e. $\sigma^{(a)}_{xy}=0$.

The vanishing Hall conductivity is more general than the specific model that we used
to derive it; see also Refs.~\cite{Sorn2024, Farajollahpour2024}. For example, in a generic tetragonal system with $D_{4h}$ symmetry, and in the presence of SOC, there are three possible types of Ising-like altermagnetic order parameters that do not induce a ferromagnetic moment, corresponding to the irreducible representations $A_{1g}^{-}$, $B_{1g}^{-}$, or $B_{2g}^{-}$. One finds vanishing Hall conductivities in
all three cases. Only for
altermagnetic order parameters that transform as $A_{2g}^{-}$ or $E_{g}^{-}$
can one generate an anomalous Hall effect. However, those are the same irreducible representations according to which a ferromagnetic order parameter transforms, resulting in a weak ferromagnetism \cite{fernandes2024topological}.

\section{Elasto-Hall conductivity} \label{sec:elasto_Hall}
Although the altermagnets discussed in the previous section do not yield an AHE, they do display a non-zero elasto-Hall conductivity $\nu^{(a)}_{\alpha \beta \gamma \delta}$ introduced in Eq.~\eqref{eq:elasto_cond}. To show this explicitly, we analyze the Hall conductivity of Eq.~\eqref{eq:sigma_omega} and the Berry curvature of Eq.~\eqref{eq:BerryC} in the presence of a small applied strain. We will first formulate the perturbation theory with respect to a generic strain field and then analyze the model from Sec. \ref{sec:two_band_model} quantitatively.

\subsection{Strain correction to the Berry curvature}
\label{sec:Hall_strain_corr}

We first discuss the correction to the Berry curvature induced by external strain in a general setting. To this end we add to the Hamiltonian of Eq.~\eqref{eq:Ham_matrix} symmetry allowed strain couplings, i.e. ${\cal H}_{\bm{k}} \rightarrow {\cal H}_{\bm{k}}+{\cal H}^\epsilon_{\bm{k}}$. In the band basis of the zero strain problem we have 
\begin{equation}
 {\cal H}_{\bm{k}} = \sum_{l} \xi_{\bm{k},l} \ket{u_{\bm{k}, l}}\bra{u_{\bm{k},l}},
\end{equation}
while the strain coupling can be rather generally described as
\begin{align}
{\cal H}^\epsilon_{\bm{k}} =& \sum_{l, l'} \epsilon^{\alpha \beta} \gamma_{ll'}^{\alpha\beta}(\bm{k}) \ket{u_{\bm{k}, l}}\bra{u_{\bm{k}, l'}}.
\label{eq:strain_coupl}
\end{align}
with coupling constants $\gamma_{ll'}^{\alpha\beta}$. Recalling that the strain tensor
\begin{equation}
 \epsilon_{\alpha\beta}=\frac{1}{2}\left(\frac{\partial U_{\alpha}}{\partial x_{\beta}}+\frac{\partial U_{\beta}}{\partial x_{\alpha}}\right),
\end{equation}
is determined by the gradients of the displacement field $U_\alpha$, we can directly relate the coefficients $\gamma_{ll'}^{\alpha\beta}(\bm{k})$ in Eq.~\eqref{eq:strain_coupl} to the matrix elements
 $ g_{ll'}^{\alpha}\left(\boldsymbol{k},\boldsymbol{q}\right)=i\sum_{\beta}\gamma_{ll'}^{\alpha\beta}\left(\boldsymbol{k}\right)q_{\beta} $ of the electron-phonon coupling to acoustic sound
\begin{equation}
 H_{{\rm el-ph}}=\sum_{\boldsymbol{k}\boldsymbol{q}\in{\rm BZ}}\sum_{ll'\alpha}g_{ll'}^{\alpha}\left(\boldsymbol{k},\boldsymbol{q}\right)U_{\alpha}\left(\boldsymbol{q}\right)c_{\boldsymbol{k+}\boldsymbol{q}l}^{\dagger}c_{\boldsymbol{k}l'}.
\end{equation}

In order to determine the elasto-conductivity, we need to include corrections to the energies and wave functions to first order in strain. The first-order perturbation gives the energy and the wave function: 
\begin{align}
\xi_{\bm{k}, l}^\epsilon &= \xi_{\bm{k},l} + \epsilon^{\alpha \beta} \gamma_{ll}^{\alpha\beta}(\bm{k}), \nonumber \\
\ket{u_{\bm{k}, l}}^\epsilon &= \ket{u_{\bm{k}, l}} +  \epsilon^{\alpha \beta} \sum_{l' \neq l} \frac{\gamma_{l' l}^{\alpha\beta}(\bm{k})}{\xi_{\bm{k}, l} - \xi_{\bm{k}, l'}} \ket{u_{\bm{k}, l'}}. \label{eq:xi_epsilon}
\end{align}
This can be used to obtain the strain correction to the Berry curvature:
\begin{equation}
 \delta \Omega_{\alpha \beta}^{(l)}(\bm{k})= \Gamma_{\alpha \beta \gamma \delta}^{(l)}(\bm{k})\epsilon^{\gamma \delta}, \label{eq:Gamma_definition}
\end{equation}
with
\begin{align}
 \Gamma_{\alpha \beta \gamma \delta}^{(l)}(\bm{k})
=& - 2 \Bigg[ \frac{\partial}{\partial k_{\alpha}} \sum_{l' \neq l} \frac{ \text{Im} [ \gamma_{l'l}^{\gamma \delta}(\bm{k}) \braket{u_{\bm{k}, l} | \partial_{k_{\beta}} u_{\bm{k}, l'}}]}{\xi_{\bm{k}, l} - \xi_{\bm{k}, l'}} \nonumber \\
& - \left( \alpha \leftrightarrow \beta \right) \Bigg] \label{eq:Berry_curv_correction}.
\end{align}
Inserting this expression into the Hall conductivity finally yields for the elasto-Hall conductivity tensor
\begin{align}
\nu^{(a)}_{\alpha \beta \gamma \delta} (\omega)
=& - \frac{e^2}{\hhbar}\sum_{\bm{k} \in \text{BZ},~l} \Bigg[ \Gamma_{\alpha \beta \gamma \delta}^{(l)}(\bm{k}) \Theta(\omega - \xi_{\bm{k},l}) \nonumber \\
&- \Omega^{(l)}_{\alpha \beta }(\bm{k}) \gamma_{ll}^{\gamma \delta}(\bm{k}) \delta(\omega - \xi_{\bm{k},l}) \Bigg] \label{eq:elasto_formula}.
\end{align}
This is our key result for the intrinsic elasto-Hall conductivity in terms of the zero-strain wave functions and energies, as well as the electron-strain matrix elements. 

From these considerations, we obtain for the elasto-Hall conductivity,  in complete analogy to Eq.~\eqref{eq:Hall_cond_freq}, the expression 
\begin{equation}
 \nu^{(a)}_{\alpha \beta \gamma \delta} =\int d\omega \left(-\frac{\partial f\left(\omega\right)}{\partial \omega}\right) \nu^{(a)}_{\alpha \beta \gamma \delta} (\omega). 
 \label{eq:etast_Hall_cond_freq}
\end{equation}
The Onsager relations of the conductivity yield under exchange of the first two indices that 
\begin{equation}
 \nu_{\alpha \beta \gamma \delta}=- \nu_{ \beta \alpha \gamma \delta},
 \label{eq:Onsager}
\end{equation}
i.e. it contributes to $\nu^{(a)}_{\alpha \beta \gamma \delta}$ as the elasto-Hall response is odd in the altermagnetic order parameter. At the same time,
the symmetry of the strain tensor $\epsilon_{\gamma \delta}=\epsilon_{\delta \gamma}$ implies that 
\begin{equation}
 \nu_{\alpha \beta \gamma \delta}= \nu_{\alpha \beta \delta \gamma }
 \label{eq:Transpose}
\end{equation}
is symmetric with respect to the last two indices. In what follows we will not explicitly give the superscript $(a)$ whenever we refer to a specific matrix element, such as $\nu_{xyxy}$.

\begin{figure}[tbp]
\begin{center}
\rotatebox{0}{\includegraphics[angle=0,width=1\linewidth]{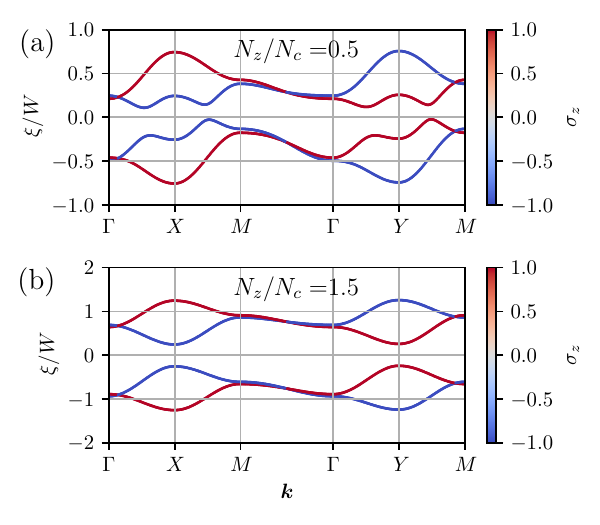}}
\caption{The band structure in the presence of applied strain for (a) $N_z = 0.5N_c$ and (b) $N_z = 1.5N_c$. The latter is shown at a larger energy scale. We set $\eta_{+} = t_2,~\eta_{-} = t_d$, $\eta_{0} = 0$ and $\epsilon_{B_{1g}} = 0.05$. The other parameters are the same as in Fig.~\ref{fig:band_no_strain}. Recall that, for $N_z > N_c$, the Dirac points of the band structure without SOC are no longer present.}
\label{fig:band_B1g}
\end{center}
\end{figure}

\subsection{Elasto-Hall conductivity of the low-energy model}
In order to gain insight on the magnitude of the effect, we next analyze the elasto-Hall conductivity of the low-energy Lieb lattice model of Sec.~\ref{sec:two_band_model}. We consider strain that transforms under $B_{1g}^+$, i.e. $\epsilon_{B_{1g}} = \epsilon_{x^2-y^2} = (\epsilon_{xx} - \epsilon_{yy})/2$. Recall that the superscripts $\pm$ for irreducible representations indicate transformation under time reversal.
According to Eq.~\eqref{eq:magnetization} this yields for the altermagnetic order parameter $N_z$ with $B_{2g}^-$ symmetry a strain-induced  magnetization that transforms under $A_{2g}^-$ i.e. that points along the $z$-direction and that should give rise to a finite value of $\nu_{xyxx}=-\nu_{yxxx}=-\nu_{xyyy}=\nu_{yxyy}$. The strain term in the Hamiltonian is given by
\begin{align}
{\cal H}_{\bm{k}}^\epsilon = &- 2\epsilon_{B_{1g}} [ - \eta_{+} (\cos k_x - \cos k_y ) \tau_{0} \nonumber \\
&+\left(\eta_{0}+ \eta_{-} (\cos k_x + \cos k_y )\right) \tau_{z}].
\end{align}
Here, $\eta_{+},~\eta_{-}$, and $\eta_{0}$ are parameters with dimension of energy that characterize the coupling to the strain field. 
We obtain the terms proportional to $\eta_{+}$ and $\eta_{-}$ by modifying the next-nearest-neighbor hopping terms along the $x$ and $y$ axes. The term $\eta_{0}$ represents the onsite energy difference between sub-lattices induced by the strain. One expects in particular $\eta_\pm$ to be of the same order as the hopping elements in Eq.~\eqref{eq:Hamiltonian}. The reason is that hopping elements between orbitals of angular momentum $\ell$ and $\ell'$ at a distance $r$ behave like $t(r)\sim t_0 (r/r_0)^{-(\ell +\ell'+1)}$\cite{Harrison2012}, such that 
\begin{equation}
\eta_\pm \sim \left. \frac{d t\left(r_0(1+\epsilon)\right)}{d\epsilon} \right|_{\epsilon=0}\sim -(\ell +\ell'+1)t_0,    
\end{equation}
with equilibrium position $r_0$. 

\begin{figure}[tbp]
\begin{center}
\rotatebox{0}{\includegraphics[angle=0,width=1\linewidth]{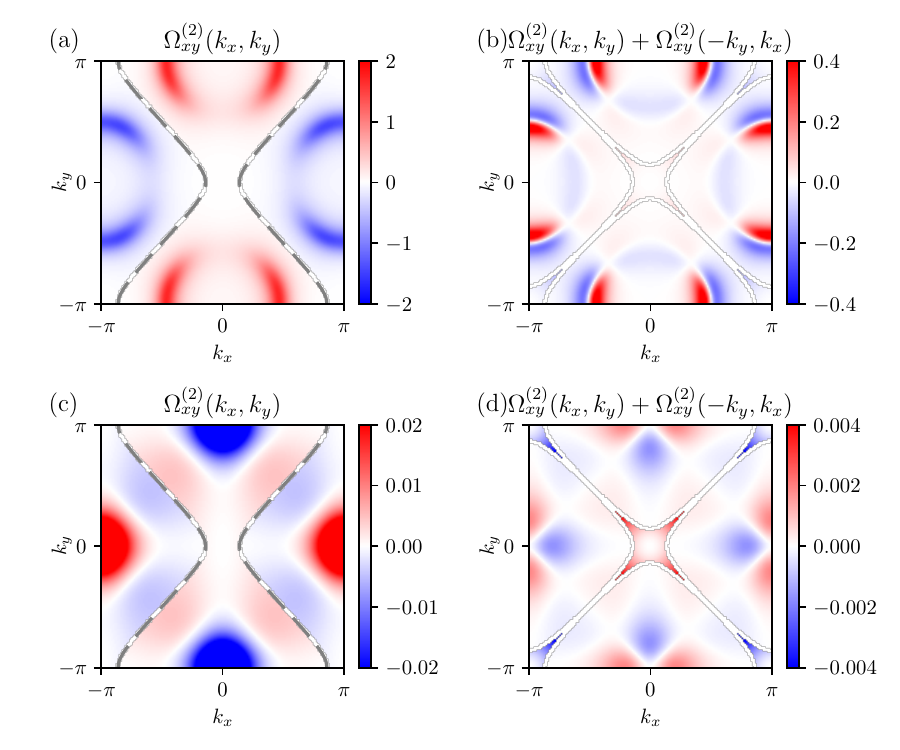}}
\caption{The Berry curvature calculated from the band structures of Fig. \ref{fig:band_B1g}(a) and (b), corresponding to an applied strain $\epsilon_{B_{1g}}=0.05$ for (a) $N_z = 0.5N_c$ and (c) $N_z = 1.5N_c$. 
The dotted lines in (a) and (c) show the nodal lines where the bands cross.
The Berry curvature superimposed by the Berry curvature rotated by $\pi/2$ around $z$-axis is shown in panels (b) and (d) for $N_z = 0.5N_c$ and $N_z = 1.5N_c$, respectively. The other parameters are the same as in Fig.~\ref{fig:band_B1g}. The order of magnitude is much smaller for panels (c) and (d), reflecting the fact that the system is topologically trivial for $N_z > N_{c}$ \cite{Antonenko2024}. Note that the superimposed pictures are shown at smaller scales as well.}
\label{fig:Berry_B1g}
\end{center}
\end{figure}

Figure~\ref{fig:band_B1g} shows the band structure with $\epsilon_{B_{1g}}=0.05$ and coupling constant values as listed in the caption. With this strain term, the degeneracy along the $\Gamma-M$ path, originally due to a mirror symmetry with respect to the $k_y = k_x$ mirror plane, is lifted, although nodal points remain, in agreement with the results of Ref. \cite{fernandes2024topological}. Furthermore, the band dispersions along $\Gamma-X$ and $\Gamma-Y$ are no longer the same. In Fig.~\ref{fig:Berry_B1g}, we show in panel (a) the Berry curvature of the second lowest band and in panel (b) the Berry curvature superimposed by the Berry curvature rotated by $\pi/2$ around $z$-axis, i.e. $\Omega_{xy}^{(2)}(k_x, k_y) + \Omega^{(2)}_{xy}(- k_y, k_x)$. The parameters used here refer to Fig. \ref{fig:band_B1g}(a), for which $N_z < N_c$. The overall structure of the Berry curvature is not altered drastically, as seen in Fig.~\ref{fig:Berry_B1g}(a). However, the distribution is skewed and no longer possesses a four-fold rotational symmetry, as demonstrated in Fig.~\ref{fig:Berry_B1g}(b). Consequently, upon integration over the first Brillouin zone of $\Omega_{xy}^{(2)}(k_x, k_y)$, a net Berry curvature monopole caused by the strain emerges. This Berry curvature monopole, in turn, causes an AHE that is linearly dependent on the strain. In Fig.~\ref{fig:Berry_B1g}(c) and (d), we show respectively $\Omega_{xy}^{(2)}(k_x, k_y)$ and $\Omega_{xy}^{(2)}(k_x, k_y) + \Omega^{(2)}_{xy}(- k_y, k_x)$ for the band structure shown in Fig. \ref{fig:band_B1g}(b), for which $N_z > N_c$. In this regime, the Dirac cones of the band dispersion without SOC are removed. Consequently, although $\Omega_{xy}^{(2)}(k_x, k_y)$ in Fig.~\ref{fig:Berry_B1g}(c) has the same form as in panel Fig.~\ref{fig:Berry_B1g}(a), the magnitude of the Berry curvature is much smaller. In Appendix~\ref{Appendix_Dirac}, we present an analytic expression for the elasto-Hall conductivity of this model in the regime $N_z<N_c$, focusing on the dominant contributions from the gapped Dirac points.   

\begin{figure}[tbp]
\begin{center}
\rotatebox{0}{\includegraphics[angle=0,width=1\linewidth]{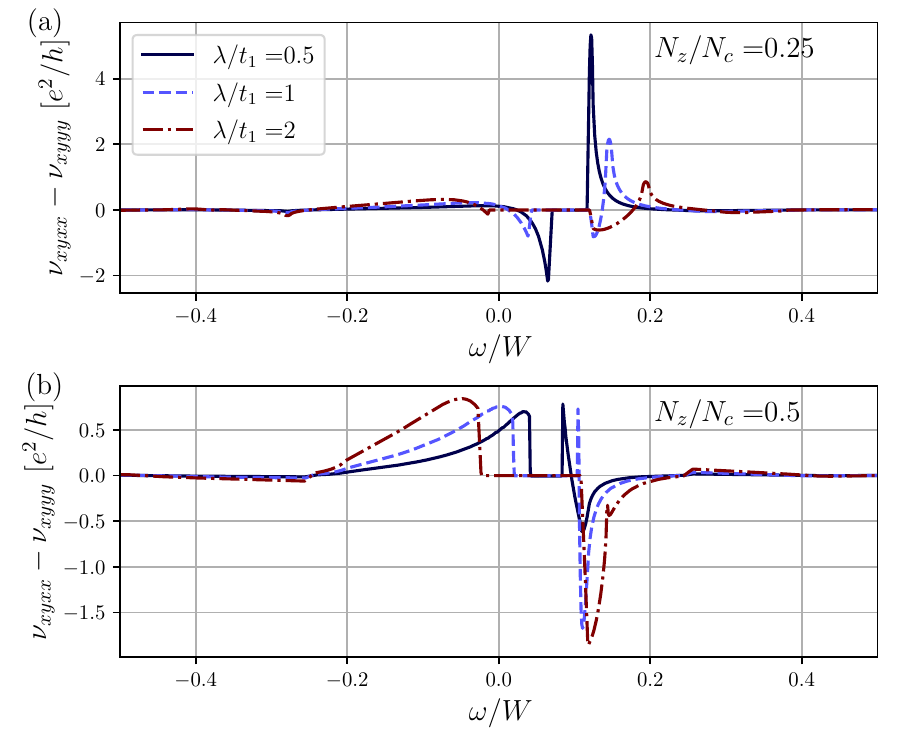}}
\caption{Elasto-Hall conductivity $\nu_{xyxx}(\omega)-\nu_{xyyy}(\omega)$ for the $d_{x^2-y^2}$ altermagnetic state described by the low-energy Lieb lattice model of Sec.~\ref{sec:two_band_model} as a function of energy $\omega$. The results are shown for (a) $N_{z}/N_{c} = 0.25$ and $N_{z}/N_{c} = 0.5$ with different values of the SOC $\lambda$. The parameters $t_{2} = \eta_{+} = 0.5t_{1}$, $t_{d} = \eta_{-} = 2t_{1}$ are the same as in Fig.~\ref{fig:band_B1g}(a). The elasto-Hall conductivity is calculated with $\epsilon_{B_{1g}}=0.05$ and obtained by dividing the finite-strain Hall conductivity by the strain value.}
\label{fig:elastoconductivity_B1g_SOC}
\end{center}
\end{figure}

\begin{figure}[tbp]
\begin{center}
\rotatebox{0}{\includegraphics[angle=0,width=1\linewidth]{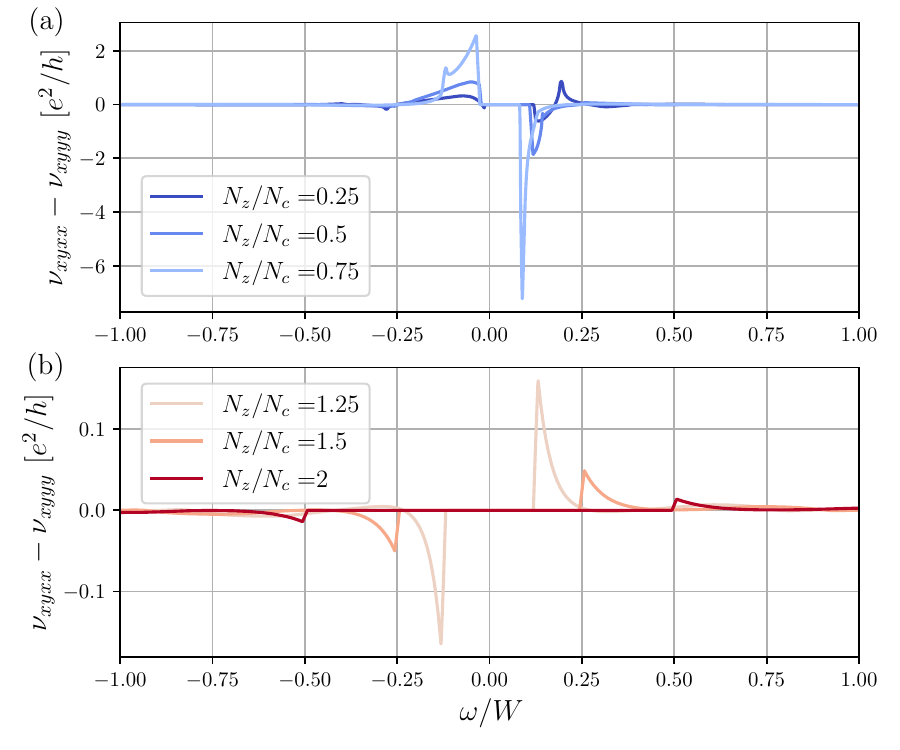}}
\caption{Elasto-Hall conductivity for different values of the altermagnetic order parameter $N_z$. 
The parameters $t_{2} = \eta_{+} = 0.5t_{1}$, $t_{d} = \eta_{-} = 2t_{1}$ are the same as in Fig.~\ref{fig:band_B1g}(a). Panels (a) and (b) show the results for $N_{z} < N_{c}$ and $N_{z} > N_{c}$, respectively. Note that the magnitude of the latter is substantially smaller due to the absence of Dirac points in the band structure.}
\label{fig:elastoconductivity_B1g_N}
\end{center}
\end{figure}

In Figs.~\ref{fig:elastoconductivity_B1g_SOC} and \ref{fig:elastoconductivity_B1g_N} we show the $\omega$ dependence of the elasto-Hall conductivity $\nu_{xyxx}(\omega) -\nu_{xyyy}(\omega)$ for the $d_{x^2-y^2}$ altermagnetic state described by the low-energy Lieb lattice model of Sec.~\ref{sec:two_band_model}. Figure~\ref{fig:elastoconductivity_B1g_SOC}(b) shows that the result only weakly depends on the magnitude of the SOC $\lambda$ compared to the panel (a). The difference is analyzed based on the Dirac model in Appendix~\ref{Appendix_Dirac} where we show the odd function component vanishes in the Dirac limit for the case of (a). Figure~\ref{fig:elastoconductivity_B1g_N} shows the non-trivial variation of the effect with the altermagnetic order parameter. We emphasize that a finite $\lambda$ is necessary to gap out the Dirac cones. Both signs change with $\omega$, i.e. at $T=0$ as a function of the Fermi energy, and the non-trivial change with the order parameter demonstrate that the response function is strongly enhanced if the spectrum is governed by the gapped Dirac points in the dispersion. In the optimal case, the elasto-Hall conductivity of our low-energy model is of the order of the quantum of conductance $e^2/\hhbar$. Hence, even if the applied strain fields are in the regime of $10^{-3}$, the anomalous conductance $\sigma_{xy}=\nu_{xyxx}(\epsilon_{xx}-\epsilon_{yy})$ should be detectable experimentally.



\subsection{Anomalous thermal elasto-Hall and Nernst effects} 

The formalism developed in the previous section can be readily generalized to determine the thermoelectric
and thermal analogs of the elasto-Hall conductivity. These response functions can also be measured in insulators, in contrast to the Hall conductivity, which is restricted to metals. To
this end we generalize Eq.~\eqref{eq:etast_Hall_cond_freq} and define
\begin{equation}
L_{\alpha\beta\gamma\delta}^{\left(n\right)}=\int d\omega\left(-\frac{\partial f\left(\omega\right)}{\partial\omega}\right)\left(\omega-\mu\right)^{n}\nu^{(a)}_{\alpha\beta\gamma\delta}\left(\omega\right).
\end{equation}

Note that the $T$-dependent elasto-Hall conductivity, defined
through Eq.~\eqref{eq:elasto_cond}, corresponds to $\nu^{(a)}_{\alpha \beta\gamma\delta}=L_{\alpha\beta\gamma\delta}^{\left(0\right)}$.
In addition, one can introduce the anomalous Nernst elasto-conductivity
\begin{equation}
j_{\alpha}=-\lambda^{(a)}_{\alpha\beta\gamma\delta}\partial_{\beta}T\epsilon_{\gamma\delta}, \label{eq:Nernst}
\end{equation}
with 
\begin{equation}
\lambda^{(a)}_{\alpha\beta\gamma\delta}=-L_{\alpha\beta\gamma\delta}^{\left(1\right)}/\left(eT\right),
\end{equation}
as well the anomalous thermal Hall elasto-conductivity 
\begin{equation}
j_{Q,\alpha}=-\mu^{(a)}_{\alpha\beta\gamma\delta}\partial_{\beta}T\epsilon_{\gamma\delta}, \label{eq:thermalHall}
\end{equation}
 with thermal current $j_{Q,\alpha}$ and
\begin{equation}\mu^{(a)}_{\alpha\beta\gamma\delta}=L_{\alpha\beta\gamma\delta}^{\left(2\right)}/\left(e^{2}T\right).
 \end{equation}
The symmetry selection rules for these quantities are the same
as those for $\nu^{(a)}_{a\beta\gamma\delta}$, which we will present in Table~\ref{table} in Sec. \ref{sec:symmetry}. Note that the Nernst elasto-conductivity $\lambda^{(a)}_{\alpha \beta \gamma \delta}$ gives rise to a corresponding elasto-Ettingshausen effect~\cite{Bridgman1924}, which relates the thermal current, the electric current, the magnetic field, and the strain tensor. 

\section{Piezomagnetism and  the effect of the strain-induced magnetization} \label{sec:piezomagnetism}
The application of strain induces, besides an AHE, a non-zero magnetization through the piezomagnetic coupling defined in Eq. (\ref{eq:magnetization}). One might be tempted to conclude that the AHE is a direct consequence of this induced magnetization. However, our explicit calculation on the low-energy Lieb lattice model did not include the induced magnetization, yet it still yielded a sizable AHE due to the distorted Berry curvature quadrupole characteristic of the altermagnetic phase. To further explore this issue, in this section we compute the magnetization induced by strain in the low-energy model via the piezomagnetic effect, and evaluate the additional contribution to the AHE arising from it.  

\begin{figure}[tbp]
\begin{center}
\rotatebox{0}{\includegraphics[angle=0,width=1\linewidth]{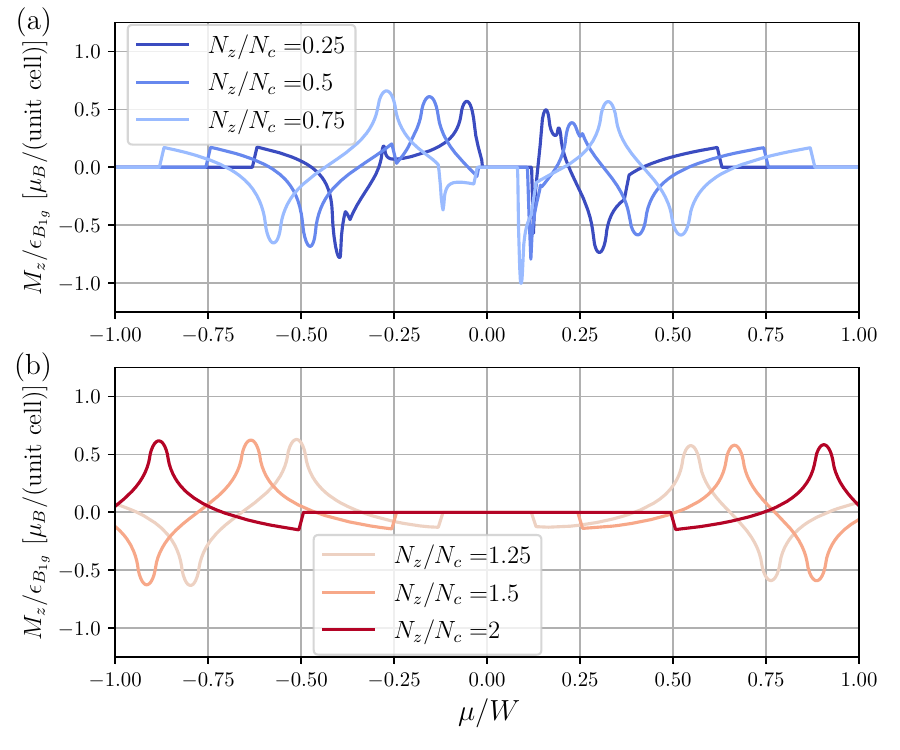}}
\caption{The net magnetization $M_z$ induced by an applied strain of $\epsilon_{B_{1g}} = 0.05$ as a function of the chemical potential $\mu$ in the low-energy model and for different values of the altermagnetic order parameter $N_z$. The upper panel corresponds to the case $N_z < N_c$ whereas the lower panel corresponds to $N_z > N_c$, where $N_c$ was introduced below Eq.~\eqref{eq:H_N}. Note that the magnetization value is divided by the strain. The other parameters are the same as in Fig.~\ref{fig:elastoconductivity_B1g_N}. }
\label{fig:mag_N_dep}
\end{center}
\end{figure}

The magnetization along the $z$ direction is given by
\begin{align}
M_z 
= \mu_{B} \sum_{\bm{k}\in \text{BZ},~l} \Theta(\mu - \xi^{\epsilon}_{\bm{k},l})\,  {}^{{\epsilon}}\hspace{-2 pt}\Braket{u_{l,\bm{k}}| \tau_{0} \sigma_{z}| u_{l,\bm{k}}}^{\epsilon},
\end{align}
where the strain-modified eigenenergies $\xi^{\epsilon}_{\bm{k},l}$ and eigenstates $\ket{u_{\bm{k}, l}}^\epsilon$ are given by Eq. (\ref{eq:xi_epsilon}).  It is therefore straightforward to compute the magnetization for a given strain value and a given chemical potential. The result, normalized by the applied strain value, is shown in Fig.~\ref{fig:mag_N_dep} for several values of the altermagnetic order parameter $N_z$. In contrast to the elasto-Hall conductivity shown in Fig. \ref{fig:elastoconductivity_B1g_N}, the piezomagnetic response is little affected by the presence or absence of Dirac points (i.e., $N_z<N_c$ or $N_z > N_c$, respectively), illustrating the different microscopic origins of the two responses.

To further compare the two response functions, we use the same perturbation approach presented in Section~\ref{sec:Hall_strain_corr} to obtain the non-zero piezomagnetic tensor of the low-energy model by calculating the linear coefficient of $M_{z}$ with respect to the strain as
\begin{align}
\left. \frac{\partial M_{z}}{\partial \epsilon_{\alpha \beta}} \right|_{\epsilon\to 0} =& \mu_{B} \sum_{\bm{k}\in \text{BZ},~l} \Bigg[ - \delta(\mu - \xi_{\bm{k},l}) \gamma^{\alpha \beta}_{ll}(\bm{k}) \Braket{u_{l,\bm{k}}| \tau_{0} \sigma_{z}| u_{l,\bm{k}}} \nonumber \\
& +\Theta(\mu - \xi_{\bm{k},l}) \sum_{l' \neq l} \frac{2 \text{Re} [\gamma^{\alpha \beta}_{l'l}(\bm{k}) \Braket{u_{l,\bm{k}}| \tau_{0} \sigma_{z}| u_{l',\bm{k}}}]}{\xi_{\bm{k},l} - \xi_{\bm{k},l'}} \Bigg]. \label{eq:piezomagnetic}
\end{align}

The first term represents the contribution due to energy change by the strain, whereas the second contribution comes from the distortion of the wave-function. However, the latter term vanishes in our model since $[H, \sigma_{z}] = 0$. Thus, we obtain:

\begin{equation}
\left. \frac{\partial M_{z}}{\partial \epsilon_{\alpha \beta}} \right|_{\epsilon\to 0} = - \mu_{B} \sum_{\bm{k}\in \text{BZ},~l} \delta(\mu - \xi_{\bm{k},l}) \gamma^{\alpha \beta}_{ll}(\bm{k}) \Braket{u_{l,\bm{k}}| \tau_{0} \sigma_{z}| u_{l,\bm{k}}}. \label{eq:piezomagnetic_final}
\end{equation}

Comparison with the formula for the elasto-Hall conductivity in Eq. (\ref{eq:elasto_formula}) reveals the fundamentally distinct natures of the two responses, as the elasto-Hall conductivity contains contributions from both the strain-dependence of the energy (like the piezomagnetic tensor) and from the Berry curvature (absent in the piezomagnetic case). To see which of these two contributions dominate the elasto-Hall conductivity, it is useful to rewrite the Hall conductivity in Eq.~\eqref{eq:sigma_omega} as
\begin{align}
\sigma^{(a)}_{xy}(\omega) =& - \frac{e^2}{\hhbar} \sum_{\bm{k}\in \text{BZ},~l} \frac{1}{2}\left( \Omega_{xy}^{(l)}(\bm{k}) - \Omega_{xy}^{(l)}(R\bm{k}) \right) \Theta(\omega - \xi_{\bm{k}, l}) \nonumber \\
& - \frac{e^2}{\hhbar} \sum_{\bm{k}\in \text{BZ},~l} \frac{1}{2}\left( \Omega_{xy}^{(l)}(\bm{k}) + \Omega_{xy}^{(l)}(R\bm{k}) \right) \Theta(\omega - \xi_{\bm{k}, l}),
\label{eq:sigma_splitted}
\end{align}
where $R$ represents a $\pi/2$ rotation. The first term, denoted by $\sigma^{(\text{E})}_{xy}(\omega)$, represents precisely the contribution from the energy change due to strain, since $\Omega_{xy}^{(l)}(\bm{k}) - \Omega_{xy}^{(l)}(R\bm{k})$ transforms as the $B_{1g}$ irreducible representation. On the other hand, the second term, denoted by $\sigma^{(\text{B})}_{xy}(\omega)$, can be identified as the contribution coming from the strain dependence of the Berry curvature, since $\Omega_{xy}^{(l)}(\bm{k}) + \Omega_{xy}^{(l)}(R\bm{k})$ transforms as $A_{1g}$.

\begin{figure}[tbp]
\begin{center}
\rotatebox{0}{\includegraphics[angle=0,width=1\linewidth]{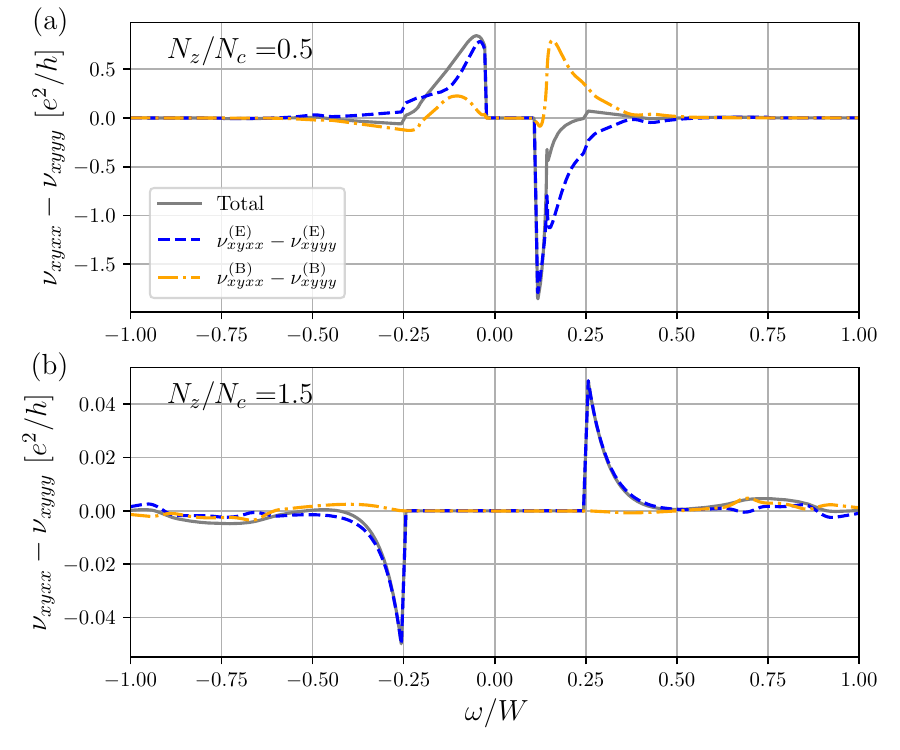}}
\caption{The total (gray) and the contribution from the energy shift (blue) and the change of the Berry curvature (orange) of elastoconductivity are shown for (a) $N_{z}/N_{c} = 0.5$ and (b) $N_{z}/N_{c} = 1.5$. }
\label{fig:elastoconductivity_N_en_shift}
\end{center}
\end{figure}

Using these expressions, we can correspondingly define two different contributions to the elasto-Hall conductivity, $\nu_{\alpha \beta \gamma \delta}^{(\text{E})}$ and $\nu_{\alpha \beta \gamma \delta}^{(\text{B})}$. 
Figure~\ref{fig:elastoconductivity_N_en_shift} shows the total elasto-Hall conductivity of our low-energy model together with its decomposition into the two components. The two components are comparable for $N_{z} < N_c$, where the electronic structure has gapped Dirac points. On the other hand, for $N_z > N_c$, the contribution due to the strain-dependence of the energy dominates. We emphasize that the contribution arising from the distortion of the Berry curvature quadrupole by the strain, $\nu_{\alpha \beta \gamma \delta}^{(\text{B})}$, has no counterpart in the expression for the piezomagnetic response, as demonstrated in Eq.~\eqref{eq:piezomagnetic}.

\begin{figure}[tbp]
\begin{center}
\rotatebox{0}{\includegraphics[angle=0,width=1\linewidth]{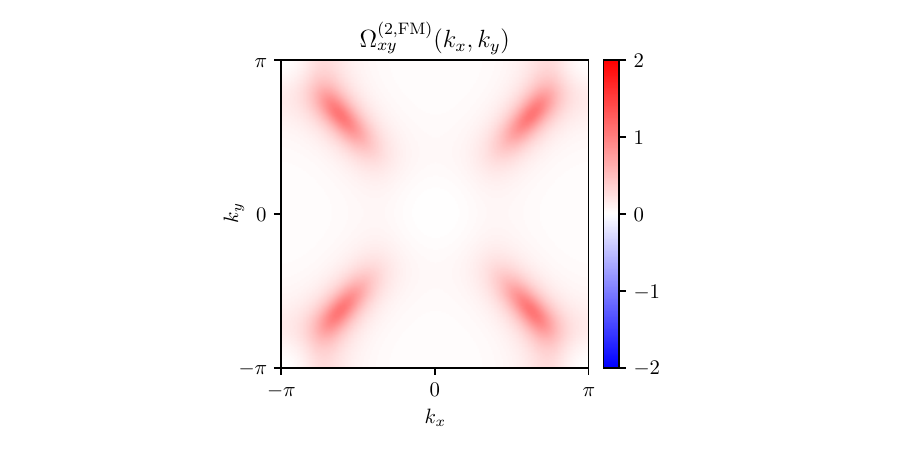}}
\caption{Berry curvature $\Omega_{xy}^{(2,\mathrm{FM})}(\bm{k})$ arising solely from the contribution of the strain-induced magnetization due to the piezomagnetic effect. Here, to isolate this contribution from that of the altermagnetic order parameter, we set $t_{2} = \eta_{+} = 0.5 t_{1}$, $t_{d} = \eta_{-} = 2t_{1}$, $N_z=0$, $J_mM_z/t_1 = 0.5$, and $\varepsilon_{B_{1g}} = 0.05$. The parameters are the same as in Fig.~\ref{fig:Berry_B1g}(a) except setting zero $N_z$ and finite $M_{z}$ in this figure.}
\label{fig:Berry_FM}
\end{center}
\end{figure}

The strain-induced magnetization $M_z$ gives an additional contribution to the AHE via its modification of the band dispersions. To calculate this additional term, we note that, in the low-energy model, $M_z$ introduces a new term in the Hamiltonian:

\begin{equation}
{\cal H}_{\boldsymbol{k}}^{M}=J_m M_z \tau_0 \sigma_z,
\end{equation}
where $J_m$ sets the Zeeman energy scale associated with the magnetization. It is straightforward to compute the Berry curvature arising due to $M_z$. To disentangle it from the Berry curvature quadrupole of the altermagnetic phase, we set $N_z = 0$ and compute $\Omega_{xy}^{(2)}(\bm{k})$. The result, shown in Fig.~\ref{fig:Berry_FM} is a Berry curvature monopole that is qualitatively different from the distorted Berry curvature quadrupole previously shown in Fig.~\ref{fig:Berry_B1g}. This confirms that the AHE of the strained altermagnet has two qualitatively different contributions: one arising from the altermagnetic order parameter, and the other from the piezomagnetic-induced magnetization.



\section{Symmetries of the elasto-Hall conductivity and its relationship with piezomagnetism} \label{sec:symmetry}
In this section, we analyze the transformation properties of the elasto-Hall conductivity tensor for a large class of relevant point groups that describe orthorhombic, tetragonal, trigonal, hexagonal, and cubic crystals. We also establish the connection between the piezomagnetic tensor $\Lambda_{\alpha \beta \gamma}$ of Eq. (\ref{eq:magnetization}) and the elasto-conductivity tensor $\nu^{(a)}_{\alpha \beta \gamma \delta}$ of Eq. (\ref{eq:elasto_cond}), and then use this relationship to derive the non-zero elasto-Hall conductivity elements.

\subsection{Analysis in terms of Jahn symbols}

It is convenient to express these tensor quantities
in terms of Jahn symbols~\cite{Jahn1949}, from which any tensor can be expressed in terms of the properties of the polar vector $V$. In this notation, $\left[\cdots\right]$ denotes symmetric
while $\left\{ \cdots\right\} $ stands for antisymmetric indices,
respectively. For example, the rank-3 tensor $A_{\alpha\beta\gamma}$ given by the Jahn symbol $\left[V^{2}\right]V$
satisfies $A_{\alpha\beta\gamma}=A_{\beta\alpha\gamma}$, while a rank-3 tensor given by
$\left\{ V^{2}\right\} V$ has $A_{\alpha\beta\gamma}=-A_{\beta\alpha\gamma}$.
Besides the polar vector $V$, the other fundamental building blocks of Jahn symbols are the rank-0 tensors $e=\pm 1$ and $a=\pm 1$, which change sign under inversion ${\cal P}$ and time reversal ${\cal T}$, respectively. Hence $a$ is crucial to ensure Onsager
reciprocity. As examples, the rank-4 elastic constant tensor corresponds to $\left[V^{2}\right]\left[V^{2}\right]$ whereas the rank-3 optical activity tensor, to $e \left[V^{2}\right]$. 

The piezomagnetic tensor $\Lambda_{\alpha \beta \gamma}$ has Jahn symbol $aeV[V^{2}]$. This can be directly obtained from Eq. (\ref{eq:magnetization}), since the magnetization transforms as a time-reversal-odd axial vector, $ae V$, and the strain tensor transforms as a symmetric rank-2 tensor $[V^{2}]$. To obtain the Jahn symbol of the elasto-Hall conductivity, we first note that the standard
conductivity can be split in a symmetric part (the ordinary
conductivity), which has Jahn symbol $[V^{2}]$, and an anti-symmetric
part (i.e. the anomalous Hall conductivity), which has Jahn symbol
$a\left\{ V^{2}\right\} $. Note that the $a$ factor is essential to ensure Onsager reciprocity, since the anti-symmetric conductivity must be odd under time-reversal. Using the fact that the strain tensor has symbol $[V^{2}]$, this means that the ordinary elasto-conductivity $\nu^{(s)}_{\alpha \beta \gamma \delta}$ has Jahn
symbol $\left[V^{2}\right]\left[V^{2}\right]$ while the elasto-Hall conductivity $\nu^{(a)}_{\alpha \beta \gamma \delta}$
has Jahn symbol $a\left\{ V^{2}\right\} \left[V^{2}\right]$. Inspection of Eqs. (\ref{eq:Nernst}) and (\ref{eq:thermalHall}) reveals that the two response functions defined by them have the same Jahn symbol $a\left\{ V^{2}\right\} \left[V^{2}\right]$. Using Jahn symbols and the Bilbao Crystallographic Server \cite{Bilbao1,Bilbao2}, we obtained the symmetry-allowed elasto-Hall conductivity elements for various relevant magnetic groups in Table \ref{table}.

The key point to relate the piezomagnetic tensor $aeV[V^{2}]$ with the elasto-Hall conductivity $a\left\{ V^{2}\right\} \left[V^{2}\right]$ is to note that $\left\{ V^{2}\right\} $ 
has the same decomposition into irreducible
representations as $e V$. This is nothing but a statement that the cross product between two polar vectors transforms as an axial vector. This implies that the irreducible representations
of the elasto-Hall conductivity tensor, $a\left\{ V^{2}\right\} \left[V^{2}\right]$,
must match those of the piezomagnetic tensor, $aeV\left[V^{2}\right]$. 
This general result, which applies to any point group, confirms the direct connection between these two observables, and allows us to identify $\nu^{(a)}_{\alpha_1 \alpha_2 \beta \gamma}$ with $\Lambda_{\alpha_3 \beta \gamma}$ provided that the indices $( \alpha_1 \alpha_2 \alpha_3)$ are a permutation of $(xyz)$. We note that a similar relationship between these two Jahn symbols was recently pointed out in Ref. \cite{Schiff2024} in the context of the antisymmetric part of the Cotton-Moutton tensor.  

Note that the Hall current is always perpendicular to the applied electric field,
irrespective of its direction. The situation is more subtle for the
Ohmic current. In unstrained crystals of high symmetry, it is entirely
longitudinal; however, in  strained, i.e. lower-symmetry crystals,
it acquires a transverse component once the electric field is not
aligned with a principal axis. This effect should not be confused with
the elasto-Hall conductivity, as they give rise to different selection
rules. This is again due to the fact that the elasto-Hall response has Jahn symbol $a\left\{ V^{2}\right\} \left[V^{2}\right]$ while the Ohmic one
has $\left[V^{2}\right]\left[V^{2}\right]$.

\begin{table*}
\begin{tabular}{|l|c|c|l|}
\hline 
 & AM irrep. & MPG & elasto-Hall conductivity\tabularnewline
\hline 
\hline  
orthorh., ${D}_{2{h}}$ ($mmm$) & $A_{1g}^{-}$& $mmm$ ($8.1.24$) & $\nu_{yzyz},$ $\nu_{zxxz}$, $\nu_{xyxy}$\tabularnewline
\hline
trigon. $D_{3d}$ ($\bar{3}m$) & $A_{1g}^-$ & $\bar{3}m$ ($20.1.71$) & $\nu_{xzxx}=\nu_{zxyy}=\nu_{zyxy}$, $\nu_{xzxz}=\nu_{yzyz}$ \tabularnewline
\hline
tetrag., ${D}_{4{h}}$ ($4/mmm$)& $A_{1g}^{-}$ & $4/mmm$ ($15.1.53$) & $\nu_{yzyz}=-\nu_{zxxz}$\tabularnewline
\hline 
 & $B_{1g}^{-}$ & $4'/mm'm$ ($15.4.56$) & $\nu_{yzyz}=\nu_{zxxz}$, $\nu_{xyxy}$\tabularnewline
\hline 
 & $B_{2g}^{-}$ &$4'/mm'm$ ($15.4.56$) & $\nu_{yzxz}=-\nu_{zxyz}$, $\nu_{xyxx}=-\nu_{xyyy}$\tabularnewline
\hline 
hexag., ${D}_{6{h}}$ ($6/mmm$)& $A_{1g}^{-}$ & $6/mmm$ ($27.1.100$)& $\nu_{yzyz}=-\nu_{zxxz}$\tabularnewline
\hline 
 & $B_{1g}^{-}$ &$6'/m'mm'$ ($27.5.104$)& $\nu_{yzxy}=\nu_{zxxx}=-\nu_{zxyy}$\tabularnewline
\hline 
 & $B_{2g}^{-}$ & $6'/m'mm'$ ($27.5.104$)& $\nu_{zxxy}=\nu_{yzyy}=-\nu_{yzxx}$\tabularnewline
\hline 
 & $E_{2g}^{-}$ & $(1,0)$ $mmm$ ($8.1.24$) & $\nu_{yzyz}$, $\nu_{zxxz}$, $\nu_{xyxy}$ \tabularnewline
 & & $(1,\frac{1}{\sqrt{3}})$ $m'm'm$ ($8.4.27$) &  $\nu_{xyxx}$, $\nu_{xyyy}$, $\nu_{xyzz}$, $\nu_{xzyz}$, $\nu_{yzxz}$ \tabularnewline & &$(1,b)$ $2/m$ ($5.1.12$) &  $\nu_{xzxx}$, $\nu_{xzyy}$,  $\nu_{xzzz}$, $\nu_{xyyz}$, $\nu_{yzyz}$, $\nu_{xzxz}$, $\nu_{xyxy}$, $\nu_{yzxy}$  \tabularnewline
\hline 
cubic, ${O}_{h}$ ($m\bar{3}m$) & $A_{1g}^{-}$& $m\bar{3}m$ ($29.1.109$) & -\tabularnewline
\hline 
 & $A_{2g}^{-}$ & $m\bar{3}m'$ ($32.4.121$) &$\nu_{yzyz}=\nu_{zxxz}=\nu_{xyxy}$\tabularnewline
\hline 
 & $E_{g}^{-}$ &$(1,0)$ $4/mmm$ ($15.1.53$) &  $\nu_{yzyz}=-\nu_{zxxz}$\tabularnewline
 & & $(1,\frac{1}{\sqrt{3}})$ $4'/mm'm$ ($15.4.56$)  &  $\nu_{yzyz}=\nu_{zxxz}$, $\nu_{xyxy}$\tabularnewline
 & &$\left(1,b\right)$ $mmm$ ($8.1.24$) & $\nu_{yzyz}$, $\nu_{zxxz}$, $\nu_{xyxy}$ \tabularnewline
\hline 
 & $T_{2g}^{-}$ & $(1,0,0)$ $4'/mm'm$ ($15.4.56$) &  $\nu_{yzxz}=-\nu_{zxyz}$, $\nu_{xyxx}=-\nu_{xyyy}$\tabularnewline
 & &$(1,1,0)$ $m'm'm$ ($8.4.27$)& $\nu_{yzxx}$, $\nu_{yzyy}$, $\nu_{yzzz}$, $\nu_{xyxz}$, $\nu_{xzxy}$ \tabularnewline
 & & $ (1,1,1)$$\bar{3}m$ ($20.1.71$)&  $\nu_{xzxy}=\nu_{zyyy}=\nu_{yzxx}$, $\nu_{xzxz}=\nu_{yzyz}$ \tabularnewline
 & & $(1,b,0)$ $2'/m'$ ($5.5.16$)&  $\nu_{xyxx}$, $\nu_{yzxx}$, $\nu_{xyyy}$, $\nu_{yzyy}$ $\nu_{xyzz}$, $\nu_{yzzz}$, $\nu_{xzyz}$, $\nu_{xyxz}$, $\nu_{yzxz}$, $\nu_{xzxy}$ \tabularnewline
 & & $(1,1,c)$ $2/m$ ($5.1.12$)& $\nu_{xzxx}$, $\nu_{xzyy}$, $\nu_{xzzz}$, $\nu_{xyyz}$, $\nu_{yzyz}$, $\nu_{xzxz}$, $\nu_{xyxy}$, $\nu_{yzxy}$\tabularnewline
 & & $(1,b,c)$ $-1$ ($2.1.3$)&  all elements $\nu_{\alpha\beta\gamma\delta}$ are nonzero where $\alpha\neq \beta$ \tabularnewline
\hline 
\end{tabular}
\caption{Non-vanishing elements of the elasto-Hall conductivity tensor $\nu^{(a)}_{\alpha \beta \gamma \delta}$ defined in Eq.~\eqref{eq:elasto_cond}. Results are shown for all pure altermagnetic order parameters (i.e., those that are not allowed by symmetry to induce a ferromagnetic moment even in the presence of SOC) in selected point groups of the orthorhombic (${D}_{2{h}}$), trigonal (${D}_{3{d}}$), tetragonal (${D}_{4{h}}$), hexagonal (${D}_{6{h}}$), and cubic (${O}_{{h}}$), crystal classes. The third column lists the magnetic point group (MPG) that results from the condensation of the order parameter listed in the second column.
Notice that due to the symmetry relations given in Eqs.~\eqref{eq:Onsager} and \eqref{eq:Transpose},
there are additional elements that occur from the permutation of the first two or the last two indices. For multi-component order parameters, which transform as a two- or three-dimensional irreducible representation, the allowed tensor elements depend on the relative magnitude of the order parameter components, parametrized as $(1,b)$ or $(1,b,c)$, respectively. Among the point groups analyzed, only an altermagnetic order parameter that transforms under $A_{1g}^-$ of the cubic group $O_h$ does not display a non-zero elasto-Hall conductivity. }
\label{table}
\end{table*}

\subsection{Analysis in terms of irreducible representations}
The relationship between the elasto-Hall conductivity and the piezomagnetic tensors discussed above can also be obtained by using the irreducible representations of point groups, as we show here. This analysis explicitly reveals the direct relationship between the allowed symmetry-elements of the elasto-Hall conductivity and of the piezomagnetic tensor with the altermagnetic order parameter. 
Following Refs.~\cite{fernandes2024topological, Steward2024}, the terms in the altermagnetic Hamiltonian reflecting the piezomagnetic coupling are given by (see also Refs. ~\citep{patri2019unveiling,aoyama2024piezomagnetic}): 
\begin{equation}
H_{\rm{am-latt}}=-\sum_{i,a,\mu}\lambda^a_{\mu,i}H_{\mu}\int d^{3}\boldsymbol{x}\epsilon_{\Gamma_{i}^{+}}\left(\boldsymbol{x}\right)N^a\left(\boldsymbol{x}\right).
\label{eq::Field_coupling}
\end{equation}

Here, $N^a\left(\boldsymbol{x}\right)$ is the $a$ component of the altermagnetic order parameter, $H_\alpha$ is the $\alpha$ component of the magnetic field, and $\epsilon_{\Gamma_{i}^{+}}$ are the components of the strain tensor that transform as the irreducible representation (irrep) $\Gamma_i^+$ of the underlying point group. Note that, because the systems we are interested in have SOC, the altermagnetic order parameter $N^a$ must transform as a time-reversal-odd irrep of the point group, which we denote $\Gamma_N^{-}$. As before, the superscripts $\pm$ indicate transformation under time reversal. Hereafter, we will focus only on the irreps $\Gamma_N^{-}$ that are not the irreps according to which the magnetization transforms, and that are even under inversion in centrosymmetric point groups. In other words, using the notation of Ref. \cite{fernandes2024topological}, we focus on so-called pure altermagnets, which do not induce ferromagnetism even in the presence of SOC (see also \cite{Scheurer2024}). Table ~\ref{table} lists, in the second column, the corresponding irreps for the highest-symmetry point groups (listed on the first column) of the orthorhombic, trigonal, tetragonal, hexagonal, and cubic crystal systems. Note that no pure altermagnetic order parameter exists in the monoclinic and triclinic point groups. On the third column, we list the corresponding magnetic point group (MPG) that results from the condensation of the altermagnetic order parameter. In cases where the altermagnetic order parameter is multi-dimensional, we list the MPGs resulting from different combinations of components. These results were obtained using the ISOTROPY software \cite{Isotropy} and the Bilbao Crystallographic Server \cite{Bilbao1,Bilbao2}. We emphasize that, while altermagnets are formally defined in terms of spin groups, because the phenomena we are interested in (e.g., AHE) are associated with SOC, we must use magnetic point groups.     

The coupling constant $\lambda^a_{\mu,i}$ in Eq. (\ref{eq::Field_coupling}) is symmetry allowed provided that $\Gamma_N^{-}\in\Gamma_{H_{\mu}}^{-}\otimes\Gamma_{\epsilon_i}^{+}$, i.e. the product representation of the $\mu$-th component of the magnetic field $\Gamma_{H_{\mu}}^{-}$
and strain $\Gamma_{\epsilon_i}^{+}$ contains the altermagnetic order parameter
representation $\Gamma_N^{-}$.

Now, the piezomagnetic response corresponds to a non-zero magnetization induced by the presence of an applied strain field in an altermagnetic state:
\begin{equation}
 M_{\mu}= \sum_{i,a}\lambda_{\mu,i}^{a}\epsilon_{\Gamma_{i}^{+}}N^{a}.
 \label{eq:magnetization2}
\end{equation}
Expanding $\epsilon_{\Gamma_{i}^{+}}=\sum_{\gamma \delta}t_{i,\gamma \delta}\epsilon_{\gamma \delta} $ with respect to the corresponding strain components, it is straightforward to obtain
the piezomagnetic tensor $\Lambda_{\mu \gamma \delta}$ from Eq.~\eqref{eq:magnetization}
\begin{equation}
    \Lambda_{\mu \gamma \delta}=\sum_{i,a}\lambda_{\mu,i}^{a}t_{i,\gamma \delta}N^{a}.
\end{equation}


In the case of an altermagnetic order parameter $N$ that transforms according to a one-dimensional irreducible representation, time reversal can be compensated by any $\Gamma_{N}^{-}$-odd operation, i.e. a symmetry operation other than time-reversal that changes the sign of  $N$.
For a one-dimensional irrep., any symmetry operation $\mathcal{R}$ can be classified as to whether it is $\Gamma_{N}^{-}$-even or $\Gamma_{N}^{-}$-odd:
\begin{align}
[\mathcal{R}, H] &= 0~(\mathcal{R}: \Gamma_{N}^{-}~\mathrm{even}), \nonumber \\
[\mathcal{R}\mathcal{T}, H] &= 0~(\mathcal{R}: \Gamma_{N}^{-}~\mathrm{odd}),
\end{align}
where $\mathcal{T}$ is time reversal. Using Eq.~\eqref{eq:BC_inv_AM}, it holds for the action of the symmetry operation $\mathcal{R}$ on the Berry curvature that
\begin{align}
\bm{\Omega}(R\bm{k}) &= \overline{R} \bm{\Omega}(\bm{k})~(\mathcal{R}: \Gamma_{N}^{-}~\mathrm{even}),\nonumber \\
\bm{\Omega}(R\bm{k}) &= - \overline{R} \bm{\Omega}(- \bm{k})~(\mathcal{R}: \Gamma_{N}^{-}~\mathrm{odd}),
\end{align}
where $R$ and $\overline{R}$ were introduced in Eq.~\eqref{eq:BC_inv_AM}.
Further, inversion symmetry tells us that $\Omega_{\mu}(\bm{k})$ is even under $\bm{k}\rightarrow -\bm{k}$
and we find for the symmetry operation of the Berry curvature:
\begin{align}
\bm{\Omega}(R\bm{k}) &= D_{\Omega}(\mathcal{R})\bm{\Omega}(\bm{k}),
\end{align}
where $D_{\Omega}(\mathcal{R}) = D_{\Gamma_{N}^{-}}(\mathcal{R}) \overline{R}$, with $D_{\Gamma_{N}^{-}}(\mathcal{R})$ the representation of the order parameter. 
Then, we obtain the representation that determines the transformation of the strain-correction to the Berry curvature as defined in Eq. (\ref{eq:Gamma_definition})
\begin{align}
\Gamma_{\mu \alpha \beta}^{(i)}(R\bm{k})
&= (D_{\Omega}(\mathcal{R}))_{\mu\nu} (R)_{\alpha \alpha'} (R^{-1})_{\beta' \beta} \Gamma_{\nu \alpha'\beta'}^{(i)}(\bm{k}).
\end{align} 
Here, in analogy to Eq.~(\ref{eq:Berry_vector}),  $\Gamma^{(i)}$ is expressed as a tensor having three indices as $\Gamma_{\mu \alpha \beta}^{(i)} = \tfrac{1}{2} \varepsilon_{\mu \nu \rho} \Gamma_{ \nu \rho\alpha \beta}^{(i)}$ with $\varepsilon_{\mu\nu\rho}$ being the Levi-Civita symbol.  At the same time, we can decompose the strain field into irreducible representations 
\begin{align}
\sum_{\alpha \beta} \epsilon^{\alpha\beta} \Gamma_{\mu \alpha \beta}^{(i)}(\bm{k}) = \sum_{\Gamma^{+}_{\epsilon_\rho }}\sum_{l} \epsilon_{\Gamma^{+}_{\epsilon_{\rho}}, l} \Gamma_{\mu;(\Gamma^{+}_{\epsilon_{\rho}}, l)}^{(i)}(\bm{k}),
\end{align}
where $l$ indicates the $l$-th component of $\Gamma^{+}_{\epsilon_{\rho}}$ irrep. Then we can write down the transformation of the Berry curvature in terms of this decomposition as
\begin{align}
\Gamma_{\mu;( \Gamma^{+}_{\epsilon_{\rho}}, l)}^{(i)}(R\bm{k}) = (D_{\Omega}(\mathcal{R}))_{\mu\nu} (D_{\Gamma^{+}_{\epsilon_{\rho}}}(\mathcal{R}))_{ll'} \Gamma_{\nu;( \Gamma^{+}_{\epsilon_{\rho}}, l')}^{(i)}(\bm{k}),
\label{eq:one_d_irreps}
\end{align}
where $D_{\Gamma^{+}_{\epsilon_{\rho}}}(\mathcal{R})$ is a representation of $\Gamma^{+}_{\epsilon_{\rho}}$.
For each $\Gamma^{+}_{\epsilon_{\rho}}$, $\Gamma_{\mu; (\Gamma^{+}_{\epsilon_{\rho}},l)}^{(i)}(\bm{k})$ transforms as $\Gamma_{N}^{-} \otimes \Gamma_{\overline{R}}^{-} \otimes \Gamma^{+}_{\epsilon_{\rho}}$ where $\Gamma_{\overline{R}}^{-}$ is the representation of axial vectors. This analysis demonstrates that the first term in Eq.~\eqref{eq:elasto_formula} gives non-vanishing $\Gamma_{\mu; (\Gamma^{+}_{\epsilon_{\rho}},l)}^{(i)}$, whenever the corresponding element of piezomagnetic tensor is nonzero. 
A similar analysis can be performed for the second term in Eq.~\eqref{eq:elasto_formula}. Using Eq.~\eqref{eq:one_d_irreps} one can verify Table~\ref{table} for all one-dimensional representations.

While the previous considerations refer to the case where the altermagnetic order parameter has a single component, we can also deduce in a similar way the components of the elasto-Hall conductivity that are linear in a multi-component altermagnetic order parameter. To this end we exploit the fact that tensor elements of the elasto-Hall conductivity are finite whenever the piezomagnetic tensor is. To linear order in the order parameter we can use Eq.~\eqref{eq:magnetization2} and determine the symmetry-allowed components of the magnetization. If a given strain $\epsilon_{\gamma \delta}$ induces a magnetization along $M_\mu$ we know that $\nu_{\alpha \beta \gamma \delta}$  is nonzero if for the given combination of indices $\varepsilon_{\alpha \beta \mu} \neq 0$, where the $\varepsilon_{\alpha \beta \mu}$ is the Levi-Civita symbol.
Consider for instance the two-component order parameter that transforms like $E_{2g}^-$ in $D_{6h}$. Then, the piezomagnetic coupling gives rise to
\begin{eqnarray}
M_{x} & = & \lambda\left(\epsilon_{yz}N_{1}-\epsilon_{xz}N_{2}\right),\nonumber \\
M_{y} & = & \lambda\left(\epsilon_{xz}N_{1}+\epsilon_{yz}N_{2}\right),\nonumber \\
M_{z} & = & \lambda'\left(2\epsilon_{xy}N_{1}-\epsilon_{x^{2}-y^{2}}N_{2}\right).
\end{eqnarray}
An order parameter with $N_1\neq 0$ yields a magnetization along the  $y$-direction if $\epsilon_{xz}\neq 0$. This gives rise to $\nu_{zxxz}\neq0$. If instead one applies the strain $\epsilon_{yz}$ but of same magnitude, the same  magnetization now points along $x$. Hence, we  obtain the relation $\nu_{zxxz}=\nu_{yzyz}$. Performing this analysis for an order parameter  $N^a=(N_1,N_2) = (1,b)$  it follows 
\begin{align}
 \nu_{yzyz}=\nu_{zxxz}=-b\nu_{yzxz}=b\nu_{zxyz},\nonumber\\
b\nu_{xyyy}=-b\nu_{xyxx}=2\nu_{xyxy},\label{eq::D6hE2g}
\end{align}
where the first line is due to  the expressions for $M_x$ and $M_y$ with same coupling $\lambda$, while the second line follows from the expression for $M_z$.
Analogously, for the two-component order parameter which transforms like $E_g^-$ in $O_h$, we obtain:
\begin{align}
\sqrt{3}\nu_{yzyz}=-\sqrt{3}\nu_{zxxz}= -b\nu_{yzyz}=-b\nu_{zxxz}=2b\nu_{xyxy}.\label{eq::EgOh}
\end{align}
Finally for the three-component order parameter that transforms like $T_{2g}$ in $O_h$, which we parametrize as $N^a = (1,b,c)$, we find:
\begin{align}
&\nu_{zxxy}=-\nu_{xyxz}=-b\nu_{yzxy}=b\nu_{xyyz}=c\nu_{yzxz}=-c\nu_{zxyz} \nonumber \\ 
&\nu_{yzyy}=-\nu_{yzzz}=b\nu_{zxzz}=-b\nu_{zxxx}=c\nu_{xyxx}=-c\nu_{xyyy}\label{eq::T2gOh}
\end{align}
These expressions all agree with the Jahn-symbol analysis that led to  Table~\ref{table}. While piezomagnetism and elasto-Hall conductivity are microscopically distinct, the symmetry conditions on the tensors are the same. Finally, we note that more components of $\nu_{\alpha \beta \gamma \delta}$ appear in Table~\ref{table} than those listed here. The additional ones refer to terms that are higher-order in the order parameter than linear.

\section{Conclusions} \label{sec:conclusions}

In this paper, we demonstrated that the hidden order of ``pure'' altermagnetic states, i.e. states with a symmetry forbidden zero-field magnetization, can be identified through the elasto-Hall conductivity, specifically the modification of the anti-symmetric part of the conductivity in response to a small applied strain. Our explicit analysis was centered on the low-energy Lieb lattice model \cite{Antonenko2024}, which describes ordered states with $B_{2g}^-$ altermagnetic order parameter in a tetragonal system, corresponding to $d_{x^2-y^2}$-wave altermagnetism. We found that the elasto-Hall-conductivity is a direct consequence of the distortion of the Berry curvature quadrupole that characterizes $d$-wave altermagnets by strain. In our low-energy model, in the regime of parameters where the band structure displays Dirac points due to a crossing between bands of same spin but opposite sublattices, we identified significant contributions to the elasto-Hall conductivity. We also computed the piezomagnetic response in this model, by which the applied strain induces a finite magnetization. We showed that the piezomagnetic response is fundamentally different from the elasto-Hall conductivity response. Importantly, the additional contribution to the AHE coming from the induced magnetization is qualitatively different from that arising from the altermagnetic order parameter, as reflected by the momentum dependence of their corresponding Berry curvatures. Therefore, our work demonstrates that the AHE and the magnetization, while simultaneously present when strain is applied in an altermagnet, arise from different types of response, illustrating that the AHE is not a consequence of the symmetry-allowed magnetization, as shown schematically in Fig. \ref{fig:magnetisation}.   

While in the model calculation we focused on the intrinsic contribution to the elasto-Hall conductivity due to the nontrivial Berry curvature, we also extended our analysis to any point group, providing a comprehensive list of symmetry-allowed elements of the elasto-Hall conductivity tensor in Table \ref{table}. This was accomplished via a tensor calculation for point groups in terms of Jahn symbols, which does not rely on a specific microscopic mechanism for the Hall response and thus gives general selection rules for the elasto-Hall conductivity of any point group of the orthorhombic, trigonal, tetragonal, hexagonal, and cubic crystal classes (monoclinic and triclinic point groups do not support ``pure'' altermagnetic orders in the presence of SOC). Our approach is applicable to all but one symmetry-allowed ordered state, offering a robust method to characterize the nature of altermagnetic ordering experimentally. While the anomalous Hall conductivity vanishes for altermagnets with symmetry-forbidden  magnetic moments, the anomalous Hall, thermal Hall, and Nernst elasto-conductivities are symmetry-allowed and provide a means to directly identify the symmetry of such magnetic states via transport measurements.

{\it Note added:} After completion of this work, a manuscript was posted where first-principles studies demonstrated an anomalous Hall, spin Hall, and valley Hall effects in strained Nb$_2$SeTeO~\cite{Jiang2024}. Where they overlap, our results are consistent with this interesting work.

\begin{acknowledgments}
We are grateful to Daniil Antonenko, Turan Birol, Alex Levchenko, Libor \v{S}mejkal, Sopheak Sorn, J\"orn Venderbos, and Yuxuan Wang for helpful discussions. K.T. acknowledges the hospitality at the Institute for Theory of Condensed Matter at KIT, where part of this work was performed, and support by the Forefront Physics and Mathematics Program to Drive Transformation (FoPM), University of Tokyo. C.R.W.S. was supported by the German Research Foundation (DFG) project SCHM 1031/12-1. M.O. was supported by JSPS KAKENHI, Grant No. JP23K03274. R.M.F. was supported by the Air Force Office of Scientific Research under Award No. FA9550-21-1-0423. J.S. and  C.R.W.S. were supported by the German Research Foundation (DFG) TRR 288-422213477 ELASTO-Q-MAT, Project A07.
\end{acknowledgments}

\appendix

\section{Analysis for an effective Dirac model}
\label{Appendix_Dirac}
The low-energy effective model which we investigate in this paper has two Dirac points along the $X-M$ line and two along the $Y-M$ line when $|N_{z}| < N_{c}$, this corresponds to the condition for zero strain case \cite{Antonenko2024}. In this appendix, we examine the elasto-conductivity using an effective model around the Dirac points. This analysis provides complementary insight into the microscopic origin of the elasto-conductivity in this model.

\subsection{Dirac points and an effective model}
The Dirac points along  the $X-M$ line, where the spin-down bands cross for $JN_{z} > 0$ at zero SOC, are located at $(\pi, \zeta k^{\ast, \downarrow})~(\zeta = \pm)$, while the Dirac points along the $Y-M$ line, where spin-up bands cross, are located at $(\zeta k^{\ast, \uparrow}, \pi)$. $k^{\ast, \sigma}$ ($\sigma = \uparrow, \downarrow$) is given by
\begin{align}
k^{\ast,\sigma} = 2 \arccos \left( \sqrt{\frac{J N_z + \sigma \varepsilon_{B_{1g}} (4\eta_{-} - 2\eta_{0} )}{4(t_d + \sigma \varepsilon_{B_{1g}} \eta_{-})}}\right),
\end{align}
where $\sigma = 1~(\sigma = -1)$ for spin-up (down) bands. The Hamiltonian is expanded around the Dirac point with the coordinates $(k_x, k_y) = (\pi + p_{x}, \zeta k^{\ast, \downarrow} + p_{y})$ for spin-down sectors and  
$(k_x, k_y) = (\zeta k^{\ast, \uparrow} + p_{x}, \pi + p_{y})$ for spin-up sectors. Then, the expanded Hamiltonian around Dirac points reads
\begin{align}
\mathcal{H}^{ \uparrow, \zeta}_{\bm{p}}
&= \xi^{\uparrow}_{0} + \zeta a^{\uparrow} p_x \tau_{0} + \zeta b^{\uparrow} p_x \tau_{z} + c^{\uparrow} p_y \tau_{x} + \lambda^{\ast,\uparrow} \tau_{y}, \nonumber \\
\mathcal{H}^{\downarrow, \zeta}_{\bm{p}}
&= \xi^{\downarrow}_{0} + \zeta a^{\downarrow} p_y \tau_{0} - \zeta b^{\downarrow} p_y \tau_{z} + c^{\downarrow} p_x \tau_{x} - \lambda^{\ast, \downarrow} \tau_{y},
\end{align}
where $\xi^{\sigma}_{0}$ is the energy of the Dirac points including the strain but neglecting the SOC term. We have introduced the following parameters to simplify the expression:
\begin{align}
a^{\sigma} &= 2\sin k^{\ast, \sigma} (t_2 - \sigma \varepsilon_{B_{1g}} \eta_{+}),\nonumber  \\
b^{\sigma} &= 2\sin k^{\ast, \sigma} (t_d + \sigma \varepsilon_{B_{1g}} \eta_{-}), \nonumber \\
c^{\sigma} &= 2 \cos \frac{k^{\ast, \sigma}}{2} t_{1}, \nonumber \\
\lambda^{\ast, \sigma} &= \lambda \sin \frac{k^{\ast, \sigma}}{2}.
\end{align}
Note that, in the limit of $\varepsilon_{B_{1g}} \to 0$, the parameters become identical between the spin-up and spin-down sectors. From the effective Hamiltonian, we can calculate the Berry curvature around the Dirac points and find the usual monopole-like structure:
\begin{align}
\Omega^{\uparrow,\zeta (\pm)}_{xy}(\bm{p}) &= \mp \frac{b^{\uparrow}c^{\uparrow}\lambda^{\ast, \uparrow}}{2[(b^{\uparrow}p_x)^2 +(c^{\uparrow}p_y)^2 + (\lambda^{\ast, \uparrow})^2]^{3/2}}, \nonumber  \\
\Omega^{\downarrow,\zeta (\pm)}_{xy}(\bm{p}) &= \pm \frac{b^{\downarrow}c^{\downarrow}\lambda^{\ast, \downarrow}}{2[(c^{\downarrow}p_x)^2 +(b^{\downarrow}p_y)^2 + (\lambda^{\ast, \downarrow})^2]^{3/2}},
\end{align}
where the supersprict $\pm$ indicates the upper $(+)$ and lower band $(-)$.

\begin{figure}[]
\begin{center}
\rotatebox{0}{\includegraphics[angle=0,width=1\linewidth]{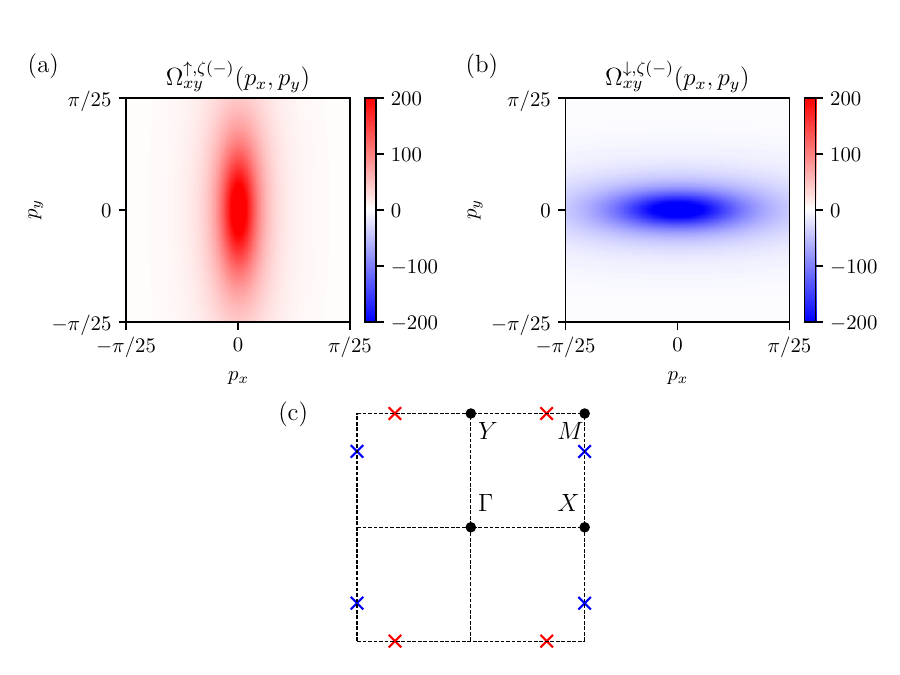}}
\caption{The Berry curvature of the lower bands of (a) spin-up section and (b) spin-down section with zero strain and $t_2 = 0.5$, $t_{d} = 2t_{1}$, $N_{z} = 0.5N_{c}$, and $\lambda/t_{1} = 0.1$. The parameters are the same as in Fig.~\ref{fig:Berry_no_strain} except for the SOC value. The locations of the Dirac points are shown in panel (c) with red (blue) colors indicating the positive (negative) sign of the Berry curvature.}
\label{fig:Berry_Dirac}
\end{center}
\end{figure}

Figure~\ref{fig:Berry_Dirac} shows the Berry curvature of the lower band for (a) spin-up band and (b) spin-down band in a small region near each Dirac point. Panel (c) indicates the locations of the Dirac points with color matching the sign of the Berry curvature. We set the SOC $\lambda/t_{1} = 0.1$ and other parameters are the same as in Fig.~\ref{fig:Berry_no_strain}. The anisotropic shape and the sign of the Berry curvature are aligned with those at each Dirac point shown in Fig.~\ref{fig:Berry_no_strain}.

\subsection{Elasto-conductivity}
\begin{figure}[]
\begin{center}
\rotatebox{0}{\includegraphics[angle=0,width=1\linewidth]{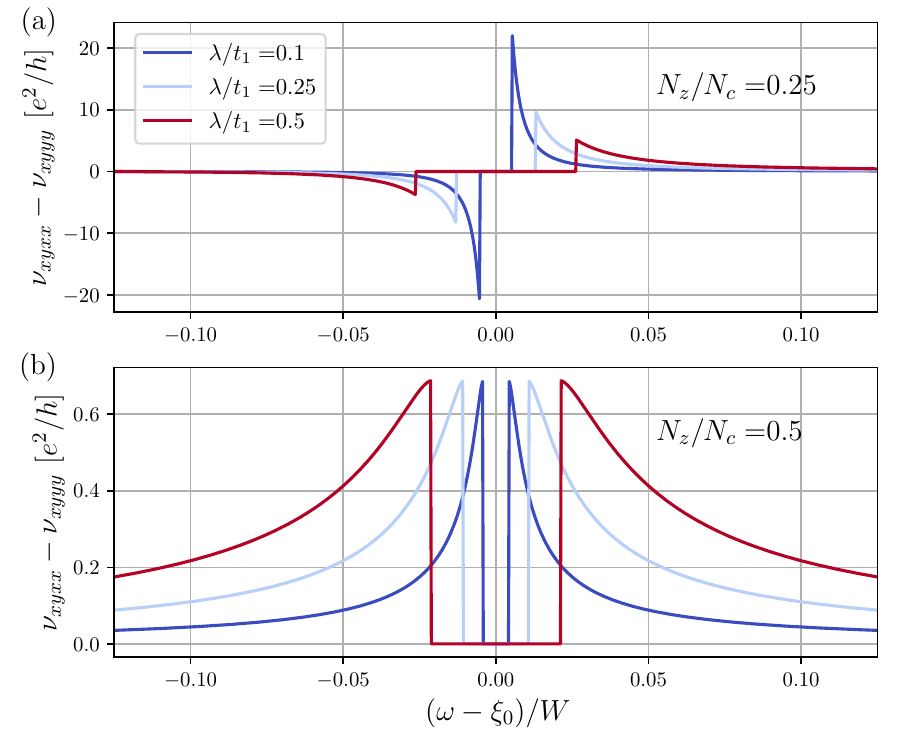}}
\caption{The elasto-Hall conductivity calculated from Eq.~\eqref{eq:elastHall_Dirac_fin} for (a) $N_{z} = 0.25N_{c}$ and (b) $N_{z} = 0.5 N_{c}$. We set $t_2 = 0.5t_{1}$, $t_{d} = 2t_{1}$, $\eta_{+} = t_{2}$, and $\eta_{-} = t_{d}$ as in Fig.~\ref{fig:elastoconductivity_B1g_N}. We plot for three SOC strengths $\lambda/t_{1} = 0.1, 0.25$, and $0.5$. }
\label{fig:elast_Dirac}
\end{center}
\end{figure}

In order to determine the elasto-Hall conductivity, we add the contributions of the AHE from each band and for finite strain. As expected, when the Fermi energy is located in the gap, each Dirac point gives a Chern number contribution  $\pm 1/2$ which cancel each other, leading to $\sigma^{(a)}_{xy}(\omega) = 0$. Hence,  we consider the case where the Fermi energy lies in the upper or lower bands. The contribution to the conductivity from the band labeled by $(\sigma, \zeta)$ is given by
\begin{align}
\sigma^{\sigma, \zeta}_{xy}(\omega) 
=& - \frac{e^2}{\hhbar} \sum_{\pm} \int \frac{d\bm{p}}{(2\pi)^2} \Omega^{\sigma, \zeta (\pm)}_{xy}(\bm{p}) \Theta(\omega - \xi^{\sigma,\zeta (\pm)}_{\bm{p}} ),
\end{align}
where $\xi^{\sigma,\zeta (\pm)}_{\bm{p}}$ is the eigen energy of the upper (lower) band. As discussed in the main text, there are two contributions to the elasto-Hall conductivity: one is the change of the Berry curvature and the other is due to a strain-induced change in the dispersion. The former is caused by the change of $b^{\sigma}$, $c^{\sigma}$, and $\lambda^{\ast, \sigma}$.
Performing the integration along the lines of
Ref.~\cite{Ogata2022}, the conductivity is:
\begin{align}
\sigma^{(a)}_{xy}(\omega) 
=& \frac{e^2}{2\pi \hhbar} \sum_{\sigma} \sigma\left( 1 - \frac{  \lambda^{\ast, \sigma}}{\sqrt{(\omega - \xi^{\sigma}_{0})^2 + (\alpha^{\sigma} \lambda^{\ast, \sigma})^2}}\right) \nonumber \\
& \times\sum_{\pm} \Theta\left( \pm( \omega  - \xi^{\sigma}_{0}) - \sqrt{1 - (\alpha^{\sigma})^2} \lambda^{\ast, \sigma} \right), 
\end{align}
where $\alpha^{\sigma} = a^{\sigma}/b^{\sigma}$ and $|\alpha^{\sigma}| < 1$ is assumed. This expression can now be used to obtain an explicit expression for the contribution of the Dirac points to the elasto-Hall conductivity:
\begin{align}
\nu^{(a)}_{xyxx}(\omega) - \nu^{(a)}_{xyyy}(\omega) =&  - \frac{e^2}{2\pi \hhbar} \cdot  \frac{ A_{+} \eta_{+} + A_{-} \eta_{-} + A_{0} \eta_{0} }{((\omega - \xi_{0})^2 + (\alpha \lambda^{\ast})^2)^{3/2}},
\label{eq:elastHall_Dirac_fin}
\end{align}
where we omitted the superscripts $\sigma$ since the coefficients become spin-independent at zero strain. 
$A_{+}$, $A_{-}$, and $A_{0}$ are given as
\begin{align}
A_{+} =&  \frac{2(\lambda^{\ast})^3t_2}{t_{d}^2} + \frac{2\lambda^{\ast}JN_{z}}{t_{d}}  (\omega - \xi_{0}) , \nonumber \\ 
A_{-}
=&  \frac{2(\lambda^{\ast})^3t_2^2}{t_{d}^3} - \frac{2\lambda^{\ast} t_{2}}{t_{d}}\left(4  - \frac{JN_{z}}{t_{d}} \right) (\omega - \xi_{0}) -\frac{\lambda^{\ast}}{t_{d}} (\omega - \xi_{0})^2 , \nonumber  \\
A_{0}
=&  \frac{4\lambda^{\ast} t_2}{t_d}(\omega - \xi_{0}) +  \frac{2\lambda^{\ast}}{4t_{d} - JN_{z}} (\omega - \xi_{0})^2 .
\end{align}
The result Eq.~\eqref{eq:elastHall_Dirac_fin} for the elasto-Hall conductivity shows that there is a large contribution from the Dirac points, in particular when the Fermi energy is near the spin-orbit induced band edge. The elasto-Hall conductivity given by Eq.~\eqref{eq:elastHall_Dirac_fin} is shown in Fig.~\ref{fig:elast_Dirac} with parameters listed in the caption. The elasto-Hall conductivity behaves differently in two cases of (a) $N_{z} = 0.25 N_{c}$ and (b) $N_{z} = 0.5 N_{c}$. This is because, in the latter case, with the chosen parameters, the odd component of the elasto-Hall conductivity with respect to $\omega - \xi_{0}$ vanishes.

\bibliography{main.bib}

 \end{document}